\documentclass[reprint, twocolumn, aps, nofootinbib, superscriptaddress, longbibliography]{revtex4-1} 
\usepackage{amsmath, amssymb}
\usepackage{multirow}
\usepackage{verbatim}
\usepackage{xfrac}
\usepackage{placeins} 

\usepackage{graphicx}
\usepackage{dcolumn}
\newcolumntype{d}[1]{D{.}{.}{#1}}
\newcommand\mc[1]{\multicolumn{1}{c}{#1}} 

\usepackage{bm}
\usepackage{hyperref}
\usepackage{url}
\usepackage{listings}
\usepackage{tabularx}
\usepackage{lipsum}
\usepackage{url}  

\usepackage{csquotes}
\usepackage{float} 
\usepackage{rotating}
\usepackage{longtable}
\usepackage{booktabs}

\AtBeginDocument{
\heavyrulewidth=.08em
\lightrulewidth=.05em
\cmidrulewidth=.03em
\belowrulesep=.65ex
\belowbottomsep=0pt
\aboverulesep=.4ex
\abovetopsep=0pt
\cmidrulesep=\doublerulesep
\cmidrulekern=.5em
\defaultaddspace=.5em
}

\lstdefinestyle{custompython}{
  belowcaptionskip=1\baselineskip,
  breaklines=true,
  frame=L,
  xleftmargin=\parindent,
  language=Python,
  showstringspaces=false,
  basicstyle=\footnotesize\ttfamily,
}

\begin{document}

\title{Subfield prestige and gender inequality in computing}

\author{Nicholas LaBerge}
    \email[]{nicholas.laberge@colorado.edu}
    \affiliation{Department of Computer Science, University of Colorado, Boulder, CO, USA}

\author{K. Hunter Wapman}
    \email[]{hunter.wapman@colorado.edu}
    \affiliation{Department of Computer Science, University of Colorado, Boulder, CO, USA}
    
 \author{Allison C. Morgan}
    \email[]{allison.morgan@colorado.edu}
    \affiliation{Department of Computer Science, University of Colorado, Boulder, CO, USA}
    
\author{Sam Zhang}
    \email[]{sam.zhang@colorado.edu}
    \affiliation{Department of Computer Science, University of Colorado, Boulder, CO, USA}

 \author{Daniel B. Larremore}
    \email[]{daniel.larremore@colorado.edu}
    \affiliation{Department of Computer Science, University of Colorado, Boulder, CO, USA}
    \affiliation{BioFrontiers Institute, University of Colorado, Boulder, CO, USA}

 \author{Aaron Clauset}
    \email[]{aaron.clauset@colorado.edu}
    \affiliation{Department of Computer Science, University of Colorado, Boulder, CO, USA}
    \affiliation{BioFrontiers Institute, University of Colorado, Boulder, CO, USA}
    \affiliation{Santa Fe Institute, Santa Fe, NM, USA}


\begin{abstract}
Women and people of color remain dramatically underrepresented among computing faculty, and improvements in demographic diversity are slow and uneven. Effective diversification strategies depend on quantifying the correlates, causes, and trends of diversity in the field. But field-level demographic changes are driven by subfield hiring dynamics because faculty searches are typically at the subfield level. Here, we quantify and forecast variations in the demographic composition of the subfields of computing using a comprehensive database of training and employment records for 6882 tenure-track faculty from 269 PhD-granting computing departments in the United States, linked with 327,969 publications. We find that subfield prestige correlates with gender inequality, such that faculty working in computing subfields with more women tend to hold positions at less prestigious institutions.  In contrast, we find no significant evidence of racial or socioeconomic differences by subfield. Tracking representation over time, we find steady progress toward gender equality in all subfields, but more prestigious subfields tend to be roughly 25 years behind the less prestigious subfields in gender representation.  These results illustrate how the choice of subfield in a faculty search can shape a department's gender diversity.
\end{abstract}


\maketitle

\section{Introduction} \label{sec:introduction}

In computing, faculty play many critical roles, including training the next generation of researchers, advancing scientific research across a diverse array of computing topics, and translating that research into practice. The composition of the academic workforce thus shapes what advances are made and who benefits from them~\cite{nielsen2017one, sugimoto2019factors, koning2021invent, kozlowski2022intersectional, evans2014attention}, in part because demographic diversity in science is known to accelerate innovation and improve problem solving \cite{hofstra2020diversity, page2008difference}. 

Despite a continued emphasis on broadening participation, women faculty in the U.S.~remain underrepresented relative to women's share of the U.S.~population by over a factor of two, and Black, Hispanic, and Native faculty by over a factor of five~\cite{computing2020taulbee, census2019quick}. Women's underrepresentation among computing researchers also persists internationally. For example, women are estimated to comprise less than 10\% of contributors to international computer science journals~\cite{mattauch2020bibliometric}. 

Explanations for this persistent pattern generally fall into two categories. On the one hand, there are generational problems, in which faculty diversity changes slowly because it takes many years for increases in diversity at the earliest stages of training to propagate up to more senior levels \cite{hargens2002demographic, marschke2007demographic}. On the other hand, there are structural and social climate problems in the U.S.~\cite{adya2005early}, in which members of underrepresented groups who aspire to or have a faculty career are pushed or pulled out of the community, which may counteract efforts to address generational problems. 
Thus, in concert, these two effects may lead to a persistent overrepresentation of majority groups~\cite{pell1996fixing, camp1997incredible, clark2005women, liu2019patching} in spite of efforts to the contrary. 

We consider a third class of problem, which exists because most faculty are hired via searches that focus on a particular subfield of computing, e.g., a search in the area of artificial intelligence. As a result, field-level demographic dynamics like gender, racial, and socioeconomic representation are in fact driven by diversity differences across computing's subfields \cite{expanding2019bizot}, and the representation of those subfields among the suppliers of future faculty \cite{clauset2015systematic}. For example, faculty searches in subfields with fewer women than other subfields are less likely to increase a department's gender diversity. Similarly, if more gender or racially diverse subfields are underrepresented at elite departments---the ones that produce the majority of future faculty~\cite{clauset2015systematic}---then the diversity of those subfields is unlikely to be reflected in new faculty hires. While there is some evidence that job searches that do not focus on a particular subfield can attract more diverse candidates~\cite{carlson2021toward, mervis2020cluster}, most searches in computing remain subfield specific. 

In practice, faculty hiring closely follows a prestige hierarchy, in which more prestigious departments produce a disproportionate share of all computing faculty~\cite{clauset2015systematic}, and a department's position within this hierarchy can be inferred directly from where its graduates were hired as faculty \cite{clauset2015systematic, de2018physical}. In this way, high prestige departments exert a correspondingly large influence over the field's demographics \cite{way2016gender}, and efforts to understand patterns, trends, and causes of demographic diversity in computing must account for effects of prestige. 

What are the implications of subfield structure and prestige in faculty hiring for diversity and demographic trends in computing? Here, we address this question by studying the intersections of gender, race, socioeconomic status, prestige, and subfield structure in computing. Our analysis uses a comprehensive database of training and employment records for 6882 tenure-track faculty from 269 PhD-granting computing departments in the United States, linked with 327,969 publications. We first quantify variation in gender, racial, socioeconomic, and prestige across the subfields of computing. We then develop simple forecasts of future gender diversity for the field as a whole, which account for diversification trends over time at the subfield level. We close with a discussion of the particular patterns and trends in faculty diversity we observe, how they relate to more general patterns in academia, and we highlight a few specific implications of our findings for long and short term efforts to increase demographic diversity among computing faculty.

\section{Data} \label{sec:data}
Our analysis spans 6882 tenured or tenure track faculty at U.S.~PhD-granting computing departments between 2010 and 2018, and includes faculty names, academic rank, institution, and the year and institution from which they received their PhD training. The underlying data are derived from a larger census-style dataset obtained under a Data Use Agreement with the Academic Analytics Research Center (AARC). For this study, we define the field of computing to include computer science departments and joint departments between computer science and information sciences, computer engineering, and other closely related departments~(SI~Appendix~\ref{sec:department_coverage}). 

To these basic education and employment variables, we add information on gender, race, childhood socioeconomic status, faculty subfield, and institutional prestige using a combination of institutional covariates, automated tools, detailed publication information, and a large survey of faculty, which we describe below. 

\subsection{Gender} 
We use a set of name-based tools to match faculty with the genders that are culturally associated with their names (SI~Appendix~\ref{sec:name_based_tools}). This methodology assigns only binary (woman/man) labels to faculty, even as we recognize that gender is nonbinary. This approach is a compromise due to the technical limitations of name-based gender methodologies and is not intended to reinforce the gender binary. We assess the reliability of our gender labeling methodology using self-labeled genders from a representative survey of computing faculty that we conducted in 2017. Comparing these gender labels, our name-based methodology agrees with self-identified genders 97\% of the time ($N = 985$). 

\subsection{Race and Childhood Socioeconomic Status} 
Faculty race is known only for the 608 faculty (8.8\%) who self-reported their race in our survey. Our survey question's design followed the U.S.~Office of Management and Budget's standards for collecting race data~\cite{office1997revisions}, which facilitates comparisons between the computing professoriate and aggregated U.S.~census data (SI~Appendix~\ref{sec:si_race}). We recognize that these categories are imperfect socially constructed representations of racial, ethnic, and place-of-origin identities. For example, the census category ``Asian'' is broad, and includes South Asians, Southeast Asians, and East Asians, among others, each of which themselves contain diverse groups. 

In our survey, 633 faculty (9.2\%) report the highest level of education achieved by their parents or legal guardians, which we use as a simple indicator of faculty's childhood socioeconomic status, following Ref.~\cite{morgan2021socioeconomic} (SI~Appendix~\ref{sec:survey_representativeness}). 

\subsection{Subfields} 
We assign each professor to a distribution over computing research subfields, based on their publications in the DBLP computer science bibliography. Using unique matches to faculty names, we algorithmically linked 5472 faculty (80\%) to their listed publications, leading to a set of 327,969 author-linked publications. 

Publications were then assigned to computing research areas using a topic model of paper titles. We first manually identified 35 computing research areas grouped into 8 computing subfields using domain knowledge and advice from subfield specialists (SI~Table ~\ref{tab:cs_structure}). For each research area, we algorithmically extracted a set of ``anchor words'' that are highly informative of publication topic as measured by mutual information \cite{jagarlamudi2012incorporating} (see SI~Appendix~\ref{sec:si_word_extraction}). These anchor words were then used to parameterize a topic model, guiding the clustering of publication titles to be aligned with our intended delineation of research areas \cite{gallagher2017anchored}. We then checked the topic assignments by manually verifying that the final, larger set of words the model learned to associate with each topic aligned with commonly agreed upon computing research areas (SI~Table~\ref{tab:cs_structure}), and that the assigned research areas for a set of well-known computing scientists agree with their known expertise (see SI~Fig.~\ref{fig:author_distributions}) 

\begin{table}[] 
\begin{tabular}{l *{2}{d{3.3}}} 
\toprule 
{} & \mc{$N$} & \mc{\% women} \\ 
\midrule\midrule 
Theory of Computer Science & 573.2 & 13.1 \\ 
Programming Languages & 181.1 & 14.2 \\ 
Numerical \& Scientific Computing & 478.8 & 14.5 \\ 
Systems & 1486.1 & 14.6 \\ 
Computational Learning & 950.8 & 17.9 \\ 
Software Engineering & 317.6 & 18.9 \\ 
Interdisciplinary Computing & 904.1 & 19.7 \\ 
Human-Computer Interaction & 580.4 & 20.0 \\ 
\midrule 
All of Computing (total) & 5472.0 & 16.7 \\ 
\midrule 
Computer Science PhDs (effective) & - & 19.0 \\ 
U.S Population (effective) & - & 51.1 \\ 
\bottomrule 
\end{tabular} 
\caption{Number of tenured or tenure-track faculty and the corresponding gender compositions for 8 computing subfields, along with the gender compositions of two reference populations, the population of computing science PhDs \cite{national2019national} and the United States population \cite{hobbs2002demographic, howden2011age, census2019quick}, each adjusted for changes over time over the years that faculty were trained.} \label{tab:frac_women} 
\end{table}

While computing research can be divided into a multiplicity of fine-grained topics, faculty hiring typically takes place at a higher level. For example, departments aiming to hire in the subfield of human-computer interaction may consider applicants who specialize in any of a variety of its nested research areas. Under our taxonomy for computing research, each of the 35 identified research areas belong to exactly one of the 8 subfields: computational learning, systems, theory of computer science, numerical \& scientific computing, human-computer interaction, interdisciplinary computing, programming languages, and software engineering (SI~Table~\ref{tab:cs_structure}). 

\begin{figure*}[] 
\includegraphics[width=0.94\textwidth]{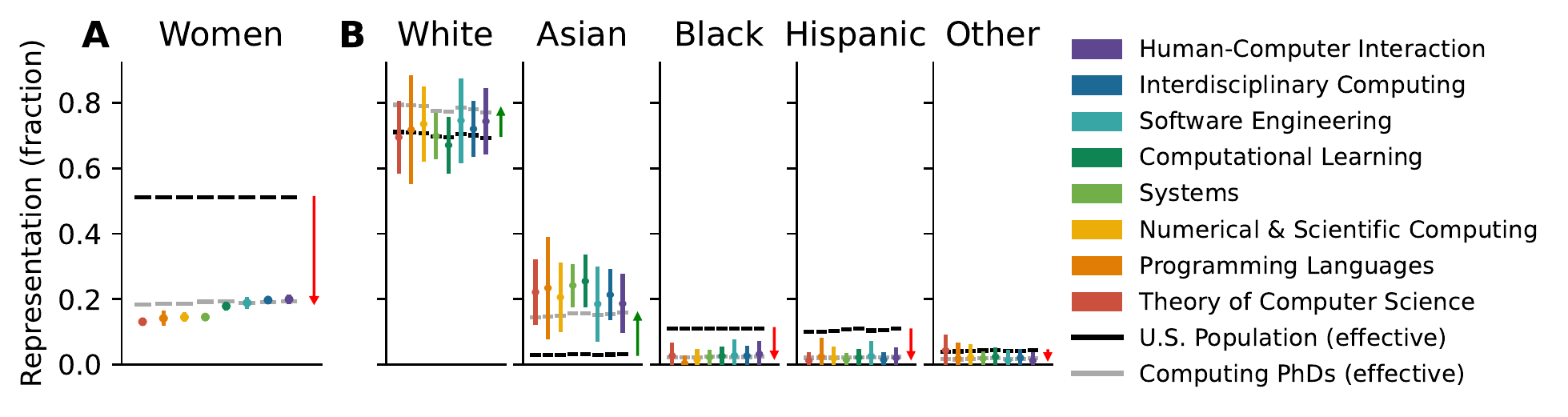} 
\caption{(A) Gender and (B) racial representation among faculty by computing subfield, with 95\% confidence intervals, along with expected levels of representation according to two time-adjusted reference populations (see text), the U.S.~population (black lines) and computing PhDs (grey lines). Gender proportions differ significantly across subfields ($\chi^{2} = 20.65$, $N = 4421$, $p~<~0.01$), but no computing subfield is representative of women in the U.S.~reference population and all but two subfields have fewer women than expected based on the reference population of computing PhDs. Racial proportions do not differ significantly across subfields ($\chi^2 = 30.88$, $N = 547$, $p = 0.67$), but do differ significantly from the U.S.~reference population. Across racial groups, representation among PhDs varies from the U.S.~population (with overrepresentation denoted by upward arrows), however, representation levels among Black, Hispanic and Other faculty are close to those expected based on their representation among recipients of computing PhDs, indicating that the largest systemic source of underrepresentation occurs prior to the transition from PhD to faculty.}\label{fig:subfields_by_race} 
\end{figure*} 

Because faculty often publish in a wide variety of areas, we assign a distribution over subfields to each professor, in proportion to the share of their publications classified into each subfield. Under this assignment, faculty belong to multiple subfields, meaning that our subsequent estimates of subfield sizes can take on non-integer values. This ``soft'' assignment scheme better captures the range of research topics that faculty work on across the boundaries of multiple subfields, compared to a ``hard'' assignment into a single subfield. We consider the hard assignment scheme in the SI~Appendix, section \ref{sec:si_assignment}.

\subsection{Institutional Prestige} 
There are many ways to quantify institutional prestige in computing, including authoritative rankings like the U.S.~News \& World Report rankings of computer science graduate departments, or the older National Research Council rankings. Such rankings have been widely criticized for their subjective selection of institutional characteristics, and for largely measuring only the inputs to the educational and research process \cite{bastedo2011how, sanoff2007us}. In contrast, publication-based approaches like that of CSRankings.org at least measure outputs of the education and research process \cite{csrankings2021}, but nevertheless depend on subjective choices and values, and is sensitive to pathologies in the academic publishing system \cite{parhami2016low, biagioli2018quality, fire2019over}. We use an alternative output-based ranking, based on institutional placement power, which quantifies prestige according to how well an institution is able to place its graduates as faculty at other institutions \cite{de2018physical} (SI~Appendix~\ref{sec:si_prestige}). This approach avoids many of the weaknesses of other measures of institutional prestige. Notably, the prestige rankings produced by this approach strongly correlate with other computer science rankings including the U.S.~News \& World Report rankings, the National Research Council (NRC) rankings, and related methods based on faculty hiring \cite{de2018physical, clauset2015systematic}, and are representative of hiring patterns across all 8 computing subfields~(SI~Fig.~\ref{fig:frac_violations}), indicating that all of these measures are capturing aspects of the underlying social processes that drive measures of prestige.

\subsection{Demographic Reference Data} 

Finally, we compare the demographic composition of current computing faculty to two reference populations: the U.S.~population and the population of U.S.~computer science PhD recipients. We reconstruct the demographics of these reference populations using the U.S.~Census \cite{gibson2002historical, humes2011overview, hobbs2002demographic, howden2011age, census2019quick} and the National Science Foundation's Survey of Earned Doctorates (SED)~\cite{sed2020national}. 

Most current computing faculty received their PhD within the past 40 years, but over that time period, the demographics of these two reference populations have changed substantially. A simple comparison of the diversity of current faculty to the diversity observed in a reference population at some particular point in time can be misleading. Instead, we construct a time-adjusted reference population, based on the demographics of the year each professor received their degree. 

For the U.S.~population, we match each professor to the U.S.~census year nearest to the year of their PhD and construct from the set of such years a weighted-average demographic distribution of the U.S. Similarly, we calculate a weighted average demographic distribution of U.S.~computing PhD recipients by matching faculty to the closest year recorded by the SED’s records of computer and information sciences doctoral recipients, which date back to 1980. While most faculty match to the survey for their exact PhD year, 11\% match to 1980, the earliest SED year, meaning they received their PhD in or prior to 1980. This procedure will tend to slightly overestimate the true diversity in the reference population. Using this methodology, we also construct reference populations for each computing subfield, which account for different age demographics across subfields. 

\section{Results} \label{sec:results}

Using these augmented data, we first quantify the gender, racial, and socioeconomic representation of faculty across computing subfields and provide a quantitative view of the demographic composition at stages prior to becoming faculty. We then ask if computing departments' choices of which subfields to hire in is predictive of overall departmental gender diversity. Then, we measure differences in subfield representation across the hierarchy of institutional prestige, and quantify how subfield prestige covaries with subfield gender diversity. Finally, we use trends in subfield diversification and growth over time to forecast the future gender diversity of the field as a whole. 

\subsection{Gender, Race, and Socioeconomic Status} 

We find wide differences in gender composition across the 8 computing subfields (Fig.~\ref{fig:subfields_by_race}A, Table \ref{tab:frac_women}; $\chi^{2} = 20.65$, $N = 4421$, $p~<~0.01$), ranging from theory of computer science (13.1\% women) and programming languages (14.2\%), to interdisciplinary computing (19.7\%) and human-computer interaction (20.0\%). No computing subfield is close to being representative of gender in the U.S.~reference population (51.1\% women). However, the proportions of women faculty in both interdisciplinary computing and human-computer interaction modestly exceed the proportion we would expect based on the time adjusted share of women PhD recipients (19.0\%). This subfield-level heterogeneity suggests that problems for gender diversity are not monolithic, and some subfields may address them more successfully than others. 

In contrast, we do not find significant differences in racial composition across subfields (Fig.~\ref{fig:subfields_by_race}B, SI~Table~\ref{tab:frac_race}; $\chi^2 = 30.88$, $N = 547$, $p = 0.67$). Rather, across all subfields, we find that some racial groups are systematically underrepresented among faculty, while others are overrepresented. To better elucidate these differences across groups, we decompose the professional pathway to becoming faculty into two stages. 

The first stage spans all steps up to and including obtaining a PhD. Hence, by comparing the proportions of different racial groups in the reference U.S.~population to those in the reference population of recipients of computing PhDs, we may quantify the relative rates of racial enrichment or depletion over this stage. Over this first stage, we find that White and Asian representation is enriched by factors of 1.1 and 5.0, respectively, while Black, Hispanic, and Native representation is depleted by factors of 4.5, 4.8, and 4.0 (Fig.~\ref{fig:subfields_by_race}B, SI~Table~\ref{tab:frac_race}). For comparison, women's representation at this stage is depleted by a factor of 2.7. 

The second stage spans all steps between obtaining the PhD and becoming faculty in a computing department. By comparing the racial proportions of the PhD recipient reference population with those of our faculty population, we can quantify the racialized rates of progression into the faculty workforce. Over this second stage, we find that White representation is depleted relative to the PhD recipients, perhaps because White PhDs are less likely to remain in academia (e.g., choosing positions in industry) or because they are less likely to receive and accept a faculty position. The enrichment of White representation in the first stage of the pathway to becoming faculty is largely compensated by their depletion in the second stage, so that White representation among computing faculty is very close to the expected levels, given the U.S.~population overall. Conversely, Asian representation is enriched in both the first and second stages, leading to a substantial overrepresentation of Asian faculty in computing relative to the U.S.~reference population. Black, Hispanic, and Native representation sees no significant enrichment or depletion in the second stage. 

This evidence suggests that the largest systematic source of racial underrepresentation occurs in the first stage of the pathway to becoming faculty, prior to the transition from PhD to faculty. This first stage includes graduate admissions and retention, which are stages known to magnify racial disparities~\cite{computing2020taulbee, mcgee2020interrogating, posselt2020equity}. We note that the data we use in this study are not equipped to determine the causes of the observed population level patterns, but observing these patterns nevertheless helps to quantify how demographics change along the professional pathway. 

\begin{figure}[b] 
\includegraphics[width=0.43\textwidth]{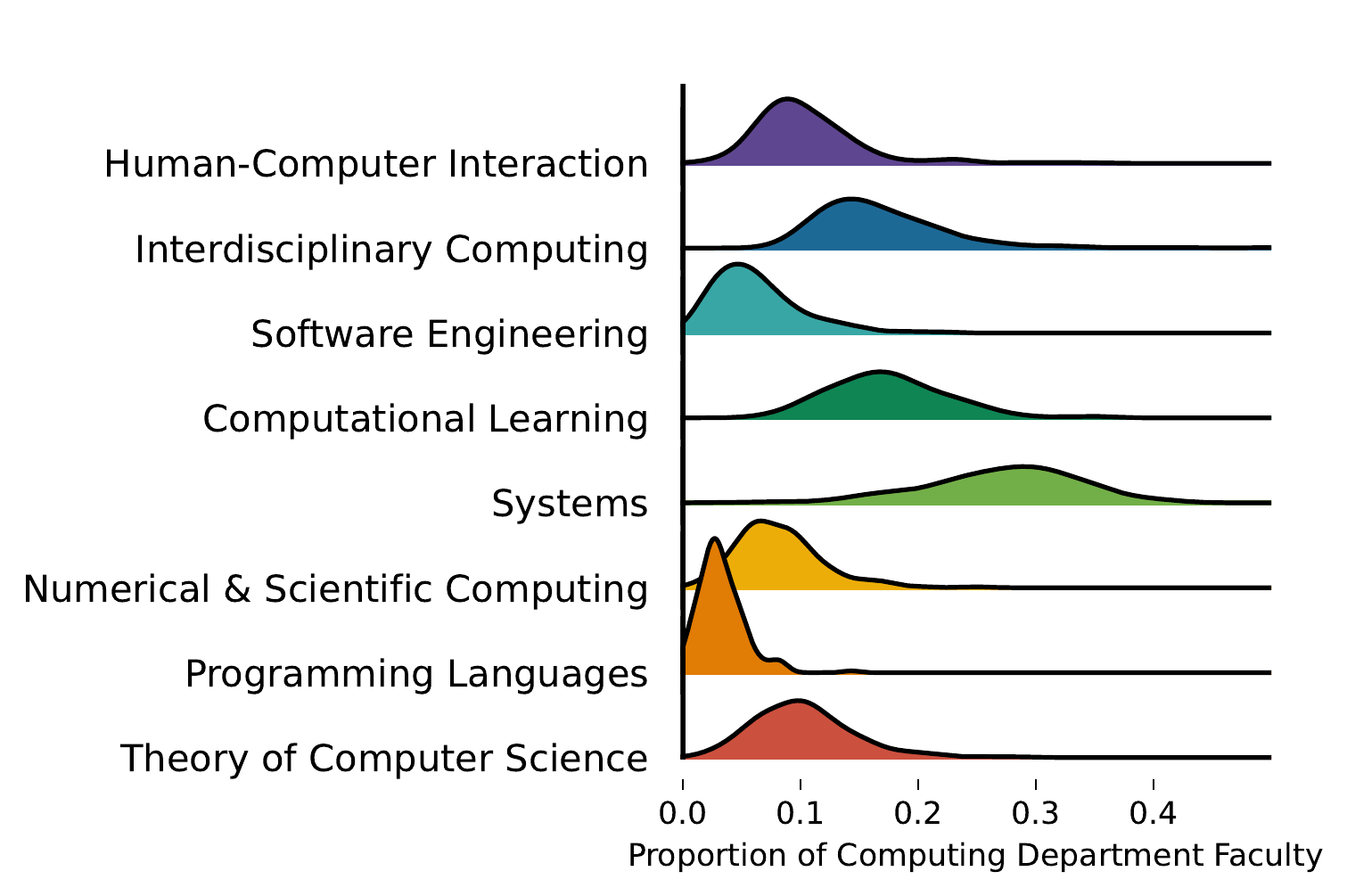} 
\caption{Distributions of subfield representation across computing departments showing that systems is the largest subfield (mean of departments 27\%), followed by computational learning (mean 17\%) while the smallest subfields are programming languages (mean 3\%) and software engineering (mean 6\%). Right skew in representation reflects a small number of departments that tend to specialize in that subfield.}\label{fig:departments_joyplot} 
\end{figure} 

\begin{figure*}[t] 
\includegraphics[width=0.29\textwidth]{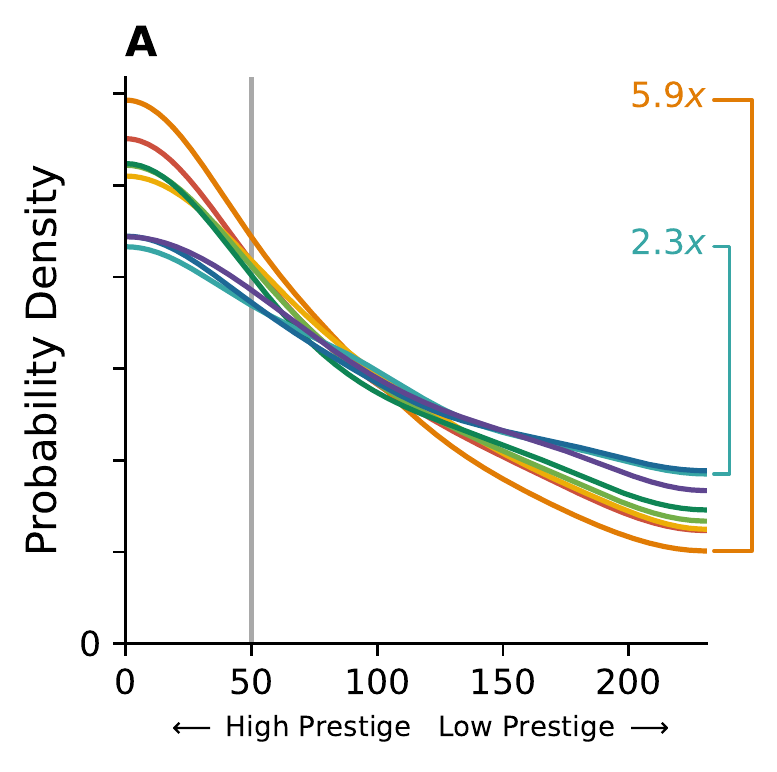} 
\includegraphics[width=0.29\textwidth]{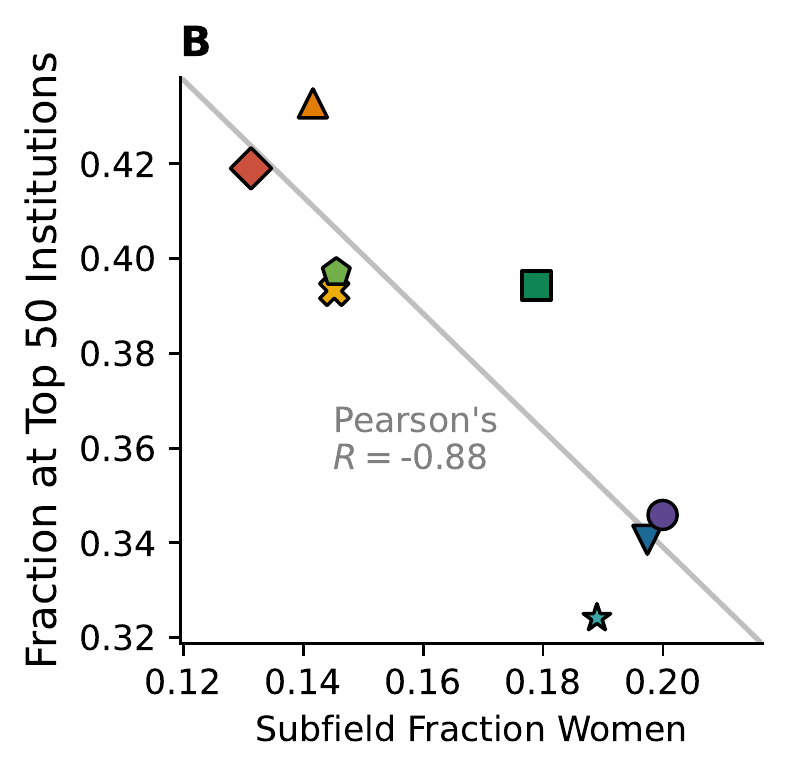} 
\includegraphics[width=0.29\textwidth]{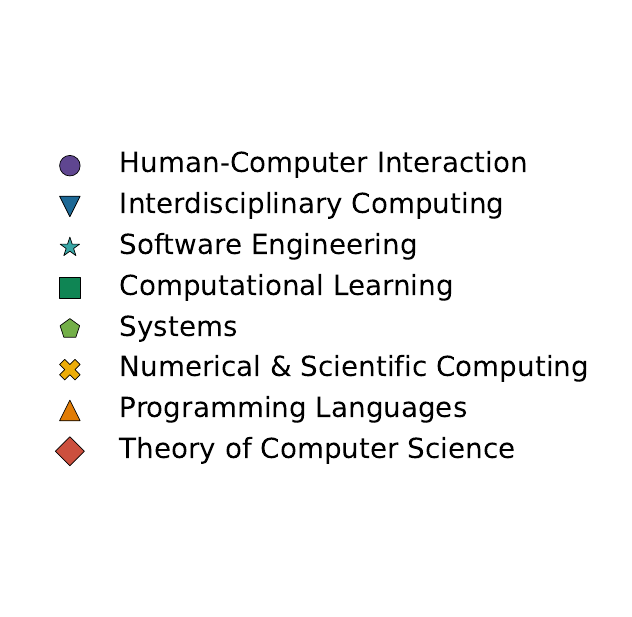} 
\caption{(A) Concentration of faculty across the computing department prestige hierarchy, for each computing subfield. Some subfields, such as programming languages and theory of computer science, have relatively high concentrations of faculty at highly prestigious departments, while others, including interdisciplinary computing, have higher faculty concentrations among less prestigious departments. Because more prestigious departments also tend to be larger, the highest concentrations occur among the high-prestige departments for all 8 subfields. To mitigate boundary effects in the kernel density estimation, the data is reflected across the minimum and maximum prestige points. (B) Fraction of faculty in each subfield working at the 50 most prestigious institutions, as a function of women's representation in that subfield. The strong negative correlation (Pearson’s $R= -0.88$, $p = 0.004$) indicates that subfields with more women tend to have a smaller share of their faculty at these elite departments.}\label{fig:prestige_histogram} 
\end{figure*} 

Past analysis found that computer science faculty tend to come from highly educated families and are between 14.5 and 28.8 times more likely to have at least one parent with a PhD than the general U.S.~population. Faculty in that category of high socioeconomic status are also more likely to hold a position at a prestigious institution: faculty at institutions ranked in the top 20\% by U.S.~News \& World Report are 57.4\% more likely to have a parent who holds a PhD than faculty at the least prestigious institutions~\cite{morgan2021socioeconomic}. Examining childhood socioeconomic status, as measured by parental educational attainment, we find no significant differences across subfields (SI~Fig.~\ref{fig:parents_edu}; $\chi^2 = 9.91$, $N = 570$, $p = 0.99$).

Faculty at the intersection of underrepresented identities are noticeably absent within our faculty sample. Black, Hispanic, and Native men comprise 3.3\% of faculty who are men, while Black, Hispanic, and Native women comprise only 0.2\% of women faculty. These small proportions preclude a detailed intersectional analysis. We return to this point in the discussion. 

\subsection{Departments} 

Most faculty are hired via searches that focus on a particular subfield of computing, e.g., a search in the area of artificial intelligence. The choice of subfield for a search may be driven by various factors, including practical needs related to the department's curriculum or strategic goals related to its research ambitions, e.g., to build on existing areas of strength or to build up a less well-established area. 

Grouping the faculty in our data by department, we find that subfields have varying representation within computing departments. The largest subfield overall focuses on systems ($N = 1486$), and also tends to include the largest share of faculty within a typical department (mean 27\%). In contrast, the smallest subfield focuses on programming languages ($N = 181$; Table~\ref{tab:frac_women}) and it, likewise, tends to include the smallest share of faculty within a typical department (mean 3\%). While most departments have some representation in each of the eight subfields, there are nevertheless departments that exhibit an unusually high degree of subfield specialization (Fig.~\ref{fig:departments_joyplot}), particularly when universities form departments dedicated to specific subfields. For instance, Carnegie Mellon University's Machine Learning department has the highest concentration of faculty in computational learning among all departments with 10 or more faculty in our dataset. Similarly, the University of Washington's Human Centered Design and Engineering department has the highest concentration faculty studying human-computer interaction. 

Some of the most gender diverse departments heavily specialize in subfields that have more women researchers. For example, the University of Washington's Human Centered Design and Engineering department and Rochester Institute of Technology's Interactive Games and Media School are among those with the highest representation of women in our dataset, and are also the most specialized in human-computer interaction and interdisciplinary computing, the two subfields with the highest proportion of women faculty (Table~\ref{tab:frac_women}). These examples highlight a connection between a department's particular subfield hiring strategy and the observed gender compositions of their faculty. In the supplement, we show that departments' subfield compositions can meaningfully improve predictions of their gender compositions~(SI~Appendix~\ref{sec:si_hiring_models}). 

\subsection{Prestige} 

The subfields of computing are correlated with prestige. On the high end, faculty in programming languages are $5.9$ times more likely to be found in the most prestigious departments than in the least prestigious departments, while on the low end, software engineering faculty are only $2.3$ times more concentrated at high prestige departments (Fig.~\ref{fig:prestige_histogram}A). 

In fact, subfield prestige correlates with subfield gender representation (Fig.~\ref{fig:prestige_histogram}B), such that more male-dominated subfields tend to have a greater share of their researchers located at higher prestige departments. In other words, men tend to be overrepresented within the more prestigious subfields, and women are more likely to be working in the less prestigious subfields. Reflecting this pattern, we find strong correlations between a field's fraction of women faculty with both the average departmental prestige for a faculty working in a given subfield (Pearson’s $R= -0.95$, $p = 0.0003$, SI~Table~\ref{tab:regression}) and the fraction of faculty in the top 50 ranked institutions (Pearson's $R = -0.88$, $p = 0.004$, Fig.~\ref{fig:prestige_histogram}B). Even after adjusting for a professor's PhD-granting institution's prestige, their publication productivity, and their gender, a multiple linear regression shows that faculty who study more male-heavy topics are still more likely to hold positions at higher prestige departments (Pearson’s $R= -0.82$, $p = 0.01$, SI~Table~\ref{tab:regression}), such that faculty fully specialized in the most prestigious subfield (programming languages) are expected to be located 12 ranks higher than faculty fully specialized in the least prestigious subfield (human-computer interaction). Because the most prestigious institutions train the majority of future faculty~\cite{clauset2015systematic, way2016gender}, the underrepresentation of gender diverse subfields among these institutions may act as a structural barrier to the gender diversity of computing as a whole.

We note that in this model, the coefficient associated with faculty gender is not statistically distinguishable from 0 ($p = 0.13$). This fact suggests that both women \emph{and men} in subfields with more women are expected to hold faculty positions lower in the prestige hierarchy. For more details on the regression findings, including regression tables, see SI~Appendix~\ref{sec:si_regression}. 

\subsection{Trends} 

Over the past 40 years, both the sizes and demographics of subfields have changed substantially. We can estimate the temporal dynamics of these variables by assigning current computing faculty to cohorts, according to the year they received their PhD, and then track how demographic and subfield distributions change over cohorts. Many faculty do not start their first faculty position until several years after completing their PhD, a pattern which would induce systematic undersampling of the most recent cohorts. To control for this effect, we report size and demographic estimates only up to the 2012 cohort. We then forecast subfield sizes and demographics 15 years into the future by extrapolating the historic trends in subfield faculty hiring over time (SI~Fig.~\ref{fig:predict_num_hires}) and the yearly gender compositions of new hires (SI Fig.~\ref{fig:predict_binomial_hiring}). 

\begin{figure*}[t]
\includegraphics[width=0.29\textwidth]{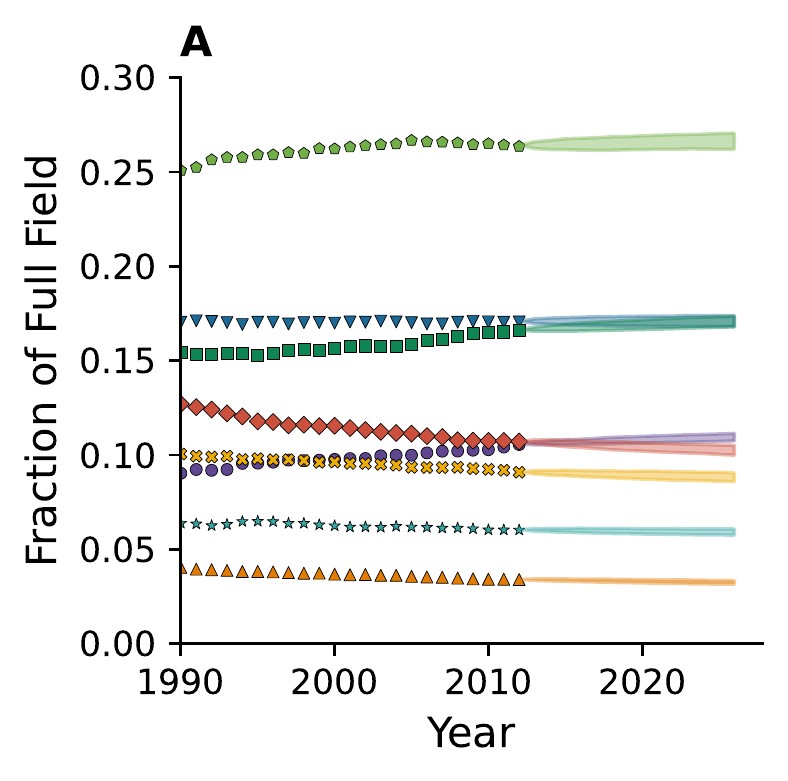} 
\includegraphics[width=0.29\textwidth]{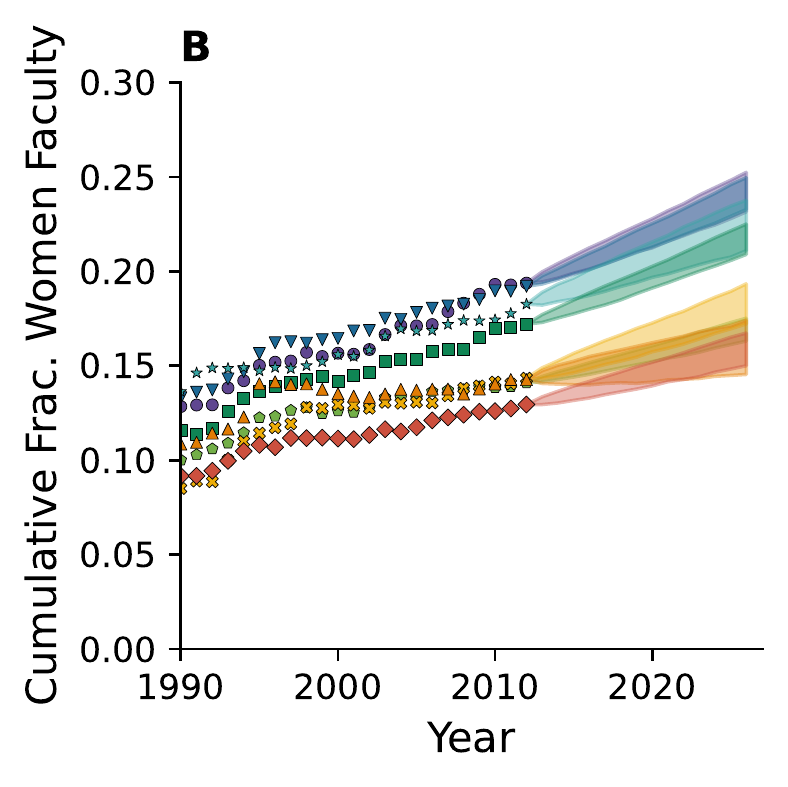} 
\includegraphics[width=0.29\textwidth]{predict_legend_binomial.pdf} 
\caption{(A) Yearly subfield size relative to computing as a whole for 1990 to 2012 and (B) cumulative fraction of faculty that are women from 1990 to 2012, for each of the eight subfields of computing, along with 95\% confidence interval forecasts, projected out to 2027, showing that we may expect the bimodal distribution of gender across subfields to continue into the foreseeable future.}\label{fig:predict} 
\end{figure*}

Analyzing these data, we find that subfields’ relative sizes have remained relatively stable over time (Fig.~\ref{fig:predict}A), even as the field as a whole has grown substantially in absolute terms. In the 22 years between 1990 and 2012, the largest increase in relative size is in human-computer interaction (+1.5\%), and the largest decline is in theory of computer science (-2.0\%). Despite enormous parallel changes in the field of computing itself since the year 2000, trends in relative subfield size appear largely stable over the past 20 years~(Fig.~\ref{fig:predict}A).

All computing subfields have increased their representation of women faculty over time, though at varying rates, and some subfields are substantially closer to parity than others (Fig.~\ref{fig:predict}B, SI~Fig.~\ref{fig:predict_binomial_hiring}). Women's representation in each annual faculty cohort has increased an average of 0.43\% per year. However, because of the generational nature of the field's composition, these annual increases in each additional cohort's gender diversity accumulate to a more modest field-level average increase of 0.2\% per year \cite{hargens2002demographic, marschke2007demographic}. These rates of change are in close agreement with past estimates for computing \cite{way2016gender, holman2018gender}. In the 22 years between 1990 and 2012, programming languages and theory of computer science have increased women's representation by the lowest relative amounts—3.4\% and 3.7\%, respectively—while interdisciplinary computing and human-computer interaction have increased by 5.9\% and 6.5\%, respectively. Although subfields’ increases in women's representation have been relatively steady over time, their slow pace produces forecasts that predict only two fields—interdisciplinary computing and human-computer interaction—are likely to reach 25\% women faculty by 2027, assuming historical trends continue. 

Our data indicate that women's representation in the four most diverse subfields (human-computer interaction, interdisciplinary computing, software engineering, and computational learning) is roughly 25 years ahead of women's representation in the four least diverse subfields, and this gap is projected to persist over the next 15 years. 

\section{Discussion} \label{sec:discussion}

Using comprehensive data on the education, employment, research subfield, and demographic variables of tenure-track faculty at U.S.-based PhD-granting computing departments, we quantified the intersection of multiple forms of demographic diversity by computing subfield, producing a detailed picture of past and likely future trends and inequalities. 

Although we find little variation in racial representation across subfields, our analysis reveals several interesting patterns about the pathway to becoming faculty, by comparing racial representation among current faculty to that of computing PhDs and the U.S.~population as a whole (Fig.~\ref{fig:subfields_by_race}B). These comparisons divide the faculty pathway into two stages. The first spans all steps prior to obtaining a PhD in computing, including doctoral training, and the second spans all steps between obtaining a PhD and becoming faculty in a computing department. Over the first stage, White and Asian representation is enriched, while Black, Hispanic, and Native representation is depleted. Over the second stage, White representation is depleted, while Asian representation is further enriched, and Black, Hispanic, and Native representation remains low without substantial change. 

These patterns are consistent with racialized factors influencing retention at multiple points in the pathway to becoming faculty in computing, and the direction and magnitude of that influence is not necessarily uniform across stages. For instance, the up and then down pattern of White representation indicates a substantial decrease in White retention after the PhD. The availability of attractive non-academic careers for computing PhDs, e.g., in the computing industry, is one plausible explanation for this decrease. White PhDs may also be more likely to pursue tenure-track jobs at non-PhD granting institutions, which are not included in our data. The down and then stable pattern of Black, Hispanic, and Native representation indicates that a large portion of the systemic effects occur prior to receiving a PhD in computing, e.g., in graduate school and college---a pattern that is well-documented by studies of race in academia \cite{ma2015race, riegle2019does}. In contrast, Asian representation in the second stage (post-PhD) increases by about the same amount that White representation decreases, suggesting additional racialized differences in achieving a faculty career after a PhD. 

Patterns underlying the underrepresentation of women computing faculty are similar to patterns of Black, Hispanic and Native underrepresentation, where the largest share of depletion occurs prior to receiving a PhD in computing (Fig.~\ref{fig:subfields_by_race}A,B). In contrast to racial diversity, we find that gender diversity varies substantially across subfields, even as overall gender diversity also remains low (16.7\%, Table \ref{tab:frac_women}), and is increasing at only about 0.2\% per year. The subfields of human-computer interaction, interdisciplinary computing, computational learning (which includes artificial intelligence), and software engineering exhibit substantially greater gender diversity among faculty (19.0\%). At the current rate of gender diversification, this level of gender diversity places them roughly 25 years ahead of the remaining four subfields (14.2\%). 

Past work has developed a number of interacting explanations for gender, racial, and intersectional underrepresentation among U.S.~computing faculty, including culturally pervasive gendered and racialized stereotypes, which may shape career decisions \cite{adya2005early, trauth2016influence, leslie2015expectations}, inhospitable educational and professional climates \cite{griffin2011re, slay2019bait, eagan2015stressing, marx2002female,shapiro2011major,witteman2019gender,clark2005women, national2018sexual}, structural disparities in education and socioeconomic status~\cite{conley2010being,edbuild2019billion,morgan2021socioeconomic,grissom2015discretion, lucas2002sociodemographic} and the unequal impact of parenthood~\cite{morgan2021unequal,hawks1998women,cech2019changing,goulden2009staying}. 
Our results do not identify any specific underlying mechanisms for differential representation, and instead quantify patterns in ways that support further research in this direction. Our results suggest that more work is needed to understand how interactions between industry and academia shape the demographic diversity of computing faculty. These interactions are likely important early in the faculty pathway, and later, e.g., where gendered or racialized hiring rates of senior faculty into industry can effectively increase the demographic diversity in academia~\cite{jurowetzki2021privatization}. 

The four most gender diverse subfields represent fully half (50.4\%, Table~\ref{tab:frac_women}) of all computing faculty. They are also substantially underrepresented among high prestige departments (Fig.~\ref{fig:prestige_histogram}A,B), which exert substantial influence over field-level norms, culture, and research agendas due to their status and their role in training the majority of computing faculty~\cite{clauset2015systematic, morgan2018prestige}. This difference holds even after controlling for factors like doctoral institution prestige, productivity, and gender itself, such that faculty working in more gender diverse subfields work at institutions, on average, 12 ranks lower than faculty working in less gender-diverse subfields (SI~Table~\ref{tab:regression}). This gender-prestige pattern illustrates a kind of systemic devaluing of women's contributions to computing overall, and the substantial size of the more diverse but less prestigious group of subfields raises the question of whether they are adequately represented among departmental curricula and degree requirements. Realigning institutional practices to reflect the true diversity of computing's subfields may help institutionalize efforts to broaden participation. 


Our retrospective analysis of subfield growth and gender shows that gender diversity is increasing at similar rates across all eight computing subfields. However, current gender diversity is essentially bimodal, with four of the eight subfields (human-computer interaction, interdisciplinary computing, software engineering, and computational learning) being substantially more gender diverse than the other four (systems, numerical \& scientific computing, programming languages, and theory of computer science). Our forecasting exercise indicates that these differences are likely to continue into the foreseeable future (Fig.~\ref{fig:predict}A,B), even as some of the less gender-diverse subfields appear to be shrinking (theory of computer science) while some of the more gender-diverse subfields are growing (computational learning, and human-computer interaction). As a result, the overall trend of slow gender diversification is highly robust to minor changes in hiring patterns among subfields. 

Our methodology for analyzing demographic patterns and trends among subfields of computing is general, and could be applied to any other academic field, given an appropriate subfield taxonomy. Applied to many fields, this approach could elucidate the systemic role that subfields play in driving field-level demographic patterns, and help identify new insights into field-specific systematic barriers to broadening participation. 

There are a number of limitations to our methodology. Although DBLP provides good general coverage of computing publications, our analysis inherits DBLP's publication inclusion bias over areas of computing, which is largest in older and in more interdisciplinary areas of research \cite{way2017misleading}. We also use the year in which a faculty received their PhD to estimate the relative sizes and gender compositions of subfields over time. This assignment assumes that faculty in our sample started their faculty jobs immediately after their PhDs and that they are representative of faculty who left jobs prior to 2010, the first year observed in our data. As a result, we are likely underestimating the historic participation of women in computing (Fig.~\ref{fig:predict}B), because women faculty have historically left their positions at higher rates than men \cite{pell1996fixing, clark2005women, cech2019changing}. This historical underestimate would imply that our estimate of gender diversification rates are likely upper bounds. Our data are limited to tenure-track faculty employed by PhD granting institutions, and do not support an analysis of contingent faculty, who make up a growing share of faculty \cite{mcnaughtan2017understanding, finley2009women} , or faculty at non-PhD granting institutions, who may exhibit different demographic compositions. We do not separately analyze faculty who hold multiple minority identities. Past research shows that people at the intersection of multiple identities often experience discrimination and exclusion beyond what would be expected from simply adding the individual elements of their identities \cite{crenshaw1989demarginalizing}. The small sample of faculty for whom we have race data limits our ability to conduct a detailed quantitative analysis of the least represented groups, and in particular, Black, Hispanic, and Native faculty, or to conduct intersectional analyses. 

We now return to the idea that explanations for slow rates of diversification in computing can be divided into categories. On the one hand, generational problems introduce a lag in faculty diversity, where, if the pathway to faculty positions were to suddenly become equitable, it would still take many years for this change to manifest as equitable representation among faculty~\cite{hargens2002demographic, marschke2007demographic}. On the other hand, there are structural and social climate problems that tend to push or pull members of underrepresented groups away from faculty positions, sometimes in different magnitudes and directions depending on the career stage \cite{adya2005early}. 

Our findings identify and quantify a third type of explanation, where the diversity of computing is driven by diversity differences across its subfields. The computing community must explore several questions before these findings can be translated into concrete policy recommendations. For example, the differences in diversity and prestige that we find across the subfield structure of computing suggest a simple departmental strategy for enhancing the probability of hiring women faculty: increase hiring in the subfields with greater gender diversity, such as human-computer interaction and interdisciplinary computing (20\% women). While this strategy may be an effective way to increase women's representation for computing as a whole, it is unlikely to reduce the heterogeneity in gender diversity across subfields. 

Future research could help shape how we design policy to increase diversity in computing, by identifying the causal mechanisms driving gender differences across subfields. On one hand, some subfields may be particularly inhospitable to women, effectively pushing women away. In this case, policy should aim to make these subfields more accessible and inclusive. On the other hand, women may, on average, be more interested in topics belonging to some subfields over others, i.e., some subfields exert stronger pulls \cite{kozlowski2022intersectional}. In this case, policy should respect the validity of women's interests by expanding the subfields that have greater pulls, instead of pushing to increase representation where there is less interest. 

An additional causal understanding of the relationship between subfield gender diversity and subfield prestige would provide further context for policy recommendations. The tendency for male-dominated areas of work to be assigned greater prestige, and hence for areas of work with greater gender diversity to be less valued, is not a phenomenon special to computing. Gendered patterns are also observed in medical subspecialties, in different areas of law \cite{buerba2020role, dixon1995stratification}, and even in less specialized positions ~\cite{blum1991between, tam1997sex}. One explanation of this pattern posits a direct causal relationship between an occupation's diversity and its prestige~\cite{england2017comparable}. If this explanation applies to computing, then it may not be feasible to simultaneously increase both a subfield's prestige and its gender diversity without first making more foundational changes to collective values and beliefs. This relationship remains untested in computing, but is an important question for diversity because the departments at the top of the prestige hierarchy tend to train the majority of future computing faculty~\cite{clauset2015systematic}. 

A subfield-focused hiring strategy alone is unlikely to increase racial or socioeconomic diversity, as we find that these faculty characteristics do not appear to correlate with subfield in our sample. Different approaches will be needed to improve representation along these dimensions, and our findings suggest these should include interventions that increase representation among PhD recipients. Some programs are available as models for future work in this direction, including the Distributed Research Experiences for Undergraduates (DREU) and the Collaborative Research Experiences for Undergraduates (CREU), two funded research programs intended to broaden participation in computing, with participants twice as likely to attend graduate school than standard REU participants~\cite{tamer2016twice}. Academic institutions are also turning to the University of Maryland, Baltimore County's Meyerhoff Scholars Program as a model for their own scholarship programs, which have been shown to markedly improve undergraduate retention and STEM graduate school matriculation for underrepresented minorities~\cite{domingo2019replicating}. Doctoral programs can additionally establish partnerships with minority serving institutions (MSIs), as modeled by the highly successful Fisk–Vanderbilt Masters-to-PhD Bridge Program, which substantially contributes to the number PhDs earned by underrepresented minorities in a number of STEM fields, but has yet to expand to computing~\cite{powell2013higher, posselt2020equity}. These are a few examples of programs that can be implemented or expanded to additonal academic institutions to increase accessibility for underrepresented groups, in conjunction with other efforts to mitigate the social climate problems in computing \cite{liu2019patching, mcgee2020interrogating, casad2021gender, stout2014now}. 


For computing departments to benefit from the innovative scientific research that diverse scientists produce \cite{hofstra2020diversity}, diversity and inclusion efforts must contend with generational, social climate, and subfield problems. For example, structural improvements to recruitment, like those suggested here, are by themselves no guarantee that diverse faculty will be adequately included and supported once they begin their faculty jobs~\cite{tienda2013diversity, smith2017diversity, slay2019bait}. Cultural change can also be slow, and also does not guarantee diverse representation among faculty. The empirical patterns and trends shown here provide new insights that can inform and support multifaceted efforts to make computing more diverse, equitable, and inclusive.

\section{Acknowledgements} \label{sec:acknowledgements}

The authors thank Bor-Yuh Evan Chang, Leysia Palen, Ben Shapiro, Huck Bennett, Joshua Grochow, and Jed Brown for helpful comments, and all survey participants for providing their valuable time. Funding: This work was supported in part by National Science Foundation Award SMA 1633791, and an Air Force Office of Scientific Research Award FA9550-19-1-0329. Competing interests: None.

\section{Citation Diversity Statement} \label{sec:diversity_statement}
Recent work in several fields of science has identified a bias in citation practices such that papers from women and other minority scholars are under-cited relative to the number of such papers in the field \cite{mitchell2013gendered,dion2018gendered,caplar2017quantitative, maliniak2013gender, Dworkin2020.01.03.894378, bertolero2021racial, wang2021gendered, chatterjee2021gender, fulvio2021imbalance}. Here we sought to proactively consider choosing references that reflect the diversity of the field in thought, form of contribution, gender, race, ethnicity, and other factors. First, we obtained the predicted gender of the first and last author of each reference by using databases that store the probability of a first name being carried by a woman \cite{Dworkin2020.01.03.894378,zhou_dale_2020_3672110}. By this measure (and excluding self-citations to the first and last authors of our current paper), our references contain 30.23\% woman(first)/woman(last), 16.23\% man/woman, 18.23\% woman/man, and 35.3\% man/man. This method is limited in that a) names, pronouns, and social media profiles used to construct the databases may not, in every case, be indicative of gender identity and b) it cannot account for intersex, non-binary, or transgender people. Second, we obtained predicted racial/ethnic category of the first and last author of each reference by databases that store the probability of a first and last name being carried by an author of color \cite{ambekar2009name, sood2018predicting}. By this measure (and excluding self-citations), our references contain 12.3\% author of color (first)/author of color(last), 14.95\% white author/author of color, 15.42\% author of color/white author, and 57.33\% white author/white author. This method is limited in that a) names and Florida Voter Data to make the predictions may not be indicative of racial/ethnic identity, and b) it cannot account for Indigenous and mixed-race authors, or those who may face differential biases due to the ambiguous racialization or ethnicization of their names. We look forward to future work that could help us to better understand how to support equitable practices in science. 

\bibliography{bib} 
\bibliographystyle{unsrt}

\section*{SUPPLEMENTARY MATERIALS} 
\renewcommand{\thefigure}{S\arabic{figure}}
\setcounter{figure}{0}
\renewcommand{\thetable}{S\arabic{table}}
\setcounter{table}{0}

\section{Coverage of Computing Departments}\label{sec:department_coverage} 
In our study, we consider faculty belonging to computer science departments and joint-departments between computer science and information sciences, computer engineering, and other closely related departments. Our sample of 269 computing departments covers 147 of the 149 (98.6\%) computer science and mixed departments that responded to the CRA's Taulbee survey for the 2017-2018 academic year \cite{computing2017taulbee}. 

\section{Name-Based Gender Tools} \label{sec:name_based_tools} 
We use a set of name-based tools to match faculty with the genders that are culturally associated with their names. For each name, we first check for consensus between two dictionary-based methods. The first uses manually annotated genders collected in a complete census of 19,000 faculty holding tenure track positions at 461 computer science, business, and history departments \cite{clauset2015systematic}, and returns the gender that is most frequently associated with a given name. The second is the python package gender-guesser \cite{genderguesser2016}, which is based on more than 40,000 names and spans multiple cultural groups. For a given name, we stop if these two methods assign the same gender. If the methods disagree, we then query the online tool Ethnea, which returns a binary gender label for names at a threshold level of prediction confidence \cite{smith2013search}. This procedure produces gender labels for 5448 (79\%) of the faculty in our dataset. We note that this methodology assigns only binary (woman/man) labels to faculty, even as we recognize that gender is nonbinary. This approach is a compromise due to the technical limitations of name-based gender methodologies and is not intended to reinforce the gender binary. The survey that we use to validate our name-based tools was approved by the University of Colorado Boulder Institutional Review Board. And, we find little evidence of systematic gender differences between groups for which genders were or were not inferred by our methodology (Pearson’s $\chi^2$ = 0.08, $N = 1139$, $p = 0.78$).

\begin{figure}[b] 
\includegraphics[width=0.45\textwidth]{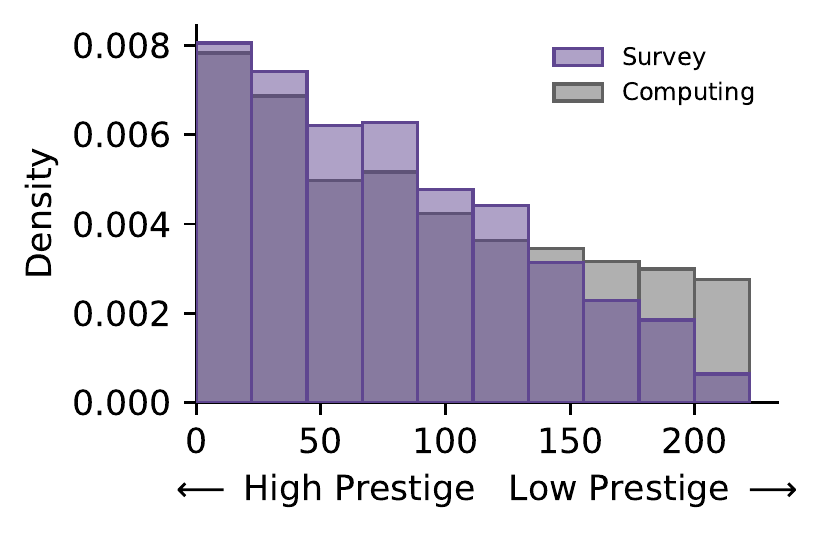} 
\caption{Prestige density histograms for the sample of computing faculty that completed our survey and for the field of computing as a whole. We find statistical evidence of prestige response bias (Kolmogorov-Smirnov~$~=~0.12$, $N = 6882$, $p < 0.01$), with faculty survey respondents skewed higher in the prestige hierarchy. To enhance interpretability, prestige is ranked ordinally according to each institution's continuous SpringRank score.}\label{fig:prestige_survey_rep} 
\end{figure}

\begin{figure}[b] 
\includegraphics[width=0.45\textwidth]{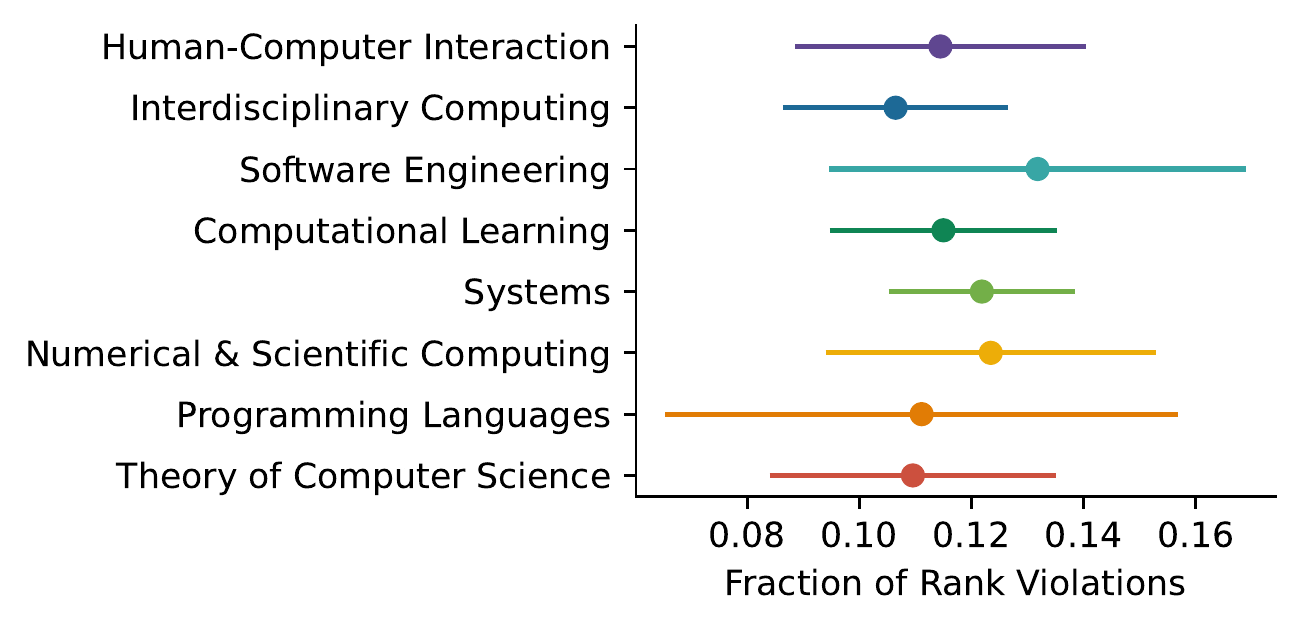} 
\caption{Fraction of faculty hiring rank violations plotted with 95\% confidence intervals for each subfield, where a rank violation occurs when a PhD trained by a lower prestige department is then hired by a higher prestige department, such that a low fraction of rank violations indicates strong adherence to the prestige hierarchy. We do not find statistical evidence of differences in the fractions of rank violations between subfields (ranging between 10.6\% violations for Interdisciplinary Computing and 13.2\% violations for Software Engineering), which suggests the full-field ranking is generally representative of hiring patterns across all subfields. 
}\label{fig:frac_violations} 
\end{figure}

\section{Operationalization of Race}\label{sec:si_race} 
In the survey, faculty selected one or more responses for their race, ethnicity, or origin from the following broad categories (i) White, (ii) Hispanic, Latino, or Spanish origin, (iii) Black or African American, (iv) Asian, (v) American Indian or Alaska Native, (vi) Native Hawaiian or other Pacific Islander, and (vii) some other race or origin. In our sample of 608 computing faculty who chose to disclose their race, ethnicity, or place of origin, 27 selected multiple responses. These faculty are divided equally among their selections, e.g., someone who selects both White and Asian will be counted as 0.5 White and 0.5 Asian. By this procedure, only 0.33 computing faculty in our sample identify as Native Hawaiian or other Pacific Islander. 

\begin{figure}[h] 
\includegraphics[width=0.45\textwidth]{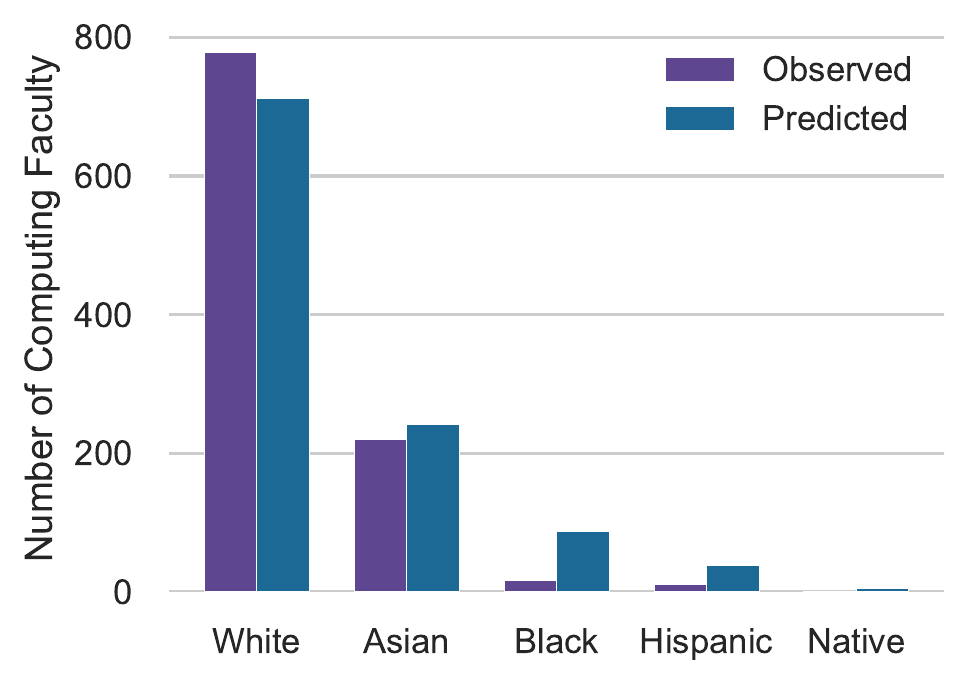} 
\caption{Here, we use race distributions over family names from the 2010 U.S.~census \cite{census2016decennial} to predict the racial demographics of computer science faculty. Rather than predicting the race of each professor individually, which has been shown to result in biased predictions, we consider each faculty to be a distribution over races as reported by the census~\cite{kozlowski2021avoiding}. For imputation in the case that faculty family names are not reported by the Census Bureau, which omits names that are reported fewer than 100 times, we use the aggregated distribution over names for which census data is available. We compare these predictions to the self-reported race data from our survey of computer science faculty ($N = 1084$) and conclude that the inferred race counts for computer science faculty do not align closely enough with self-reported race information to apply these methods in our analysis. In particular, this method overestimates the representation of Black, Hispanic, and Native faculty, by a factor of 5.5, 3.4, and 4.9, respectively.} 
\label{fig:race_obs_vs_pred} 
\end{figure}

We do not use name-based race classifiers to expand this coverage as we do for faculty gender, because we found that the tools we considered for this purpose were inaccurate with respect to faculty self-identified race even after taking measures to reduce prediction bias~\cite{kozlowski2021avoiding}, especially when predicting the representation of Black, Hispanic, and Native computing faculty (SI~Fig.~\ref{fig:race_obs_vs_pred}).

\section{Survey Representativeness}\label{sec:survey_representativeness} 
We test whether the sample of faculty who responded to our survey is representative of the complete set of computing faculty with respect to academic rank and institutional prestige, two career-relevant covariates. We find no statistically significant difference in the sample from the complete set of computing faculty by academic rank, across assistant, associate, and full professors (Pearson’s $\chi^2 = 1.58$, $N = 6882$, $p = 0.45$). However, we do find evidence that faculty survey respondents (mean ordinal rank = 77.7) are modestly more concentrated at higher prestige institutions than expected based on the complete set of computing faculty (mean ordinal rank = 91.2), indicating a modest bias toward faculty at more prestigious institutions (Kolmogorov-Smirnov~$~=~0.12$, $N = 6882$, $p < 0.01$, SI~Fig.~\ref{fig:prestige_survey_rep}).

\section{Operationalization of Prestige}\label{sec:si_prestige} 
To construct this prestige measure, we first assemble a comprehensive faculty hiring network for the 6882 computing faculty in our dataset, where nodes are institutions offering computing degrees, and each directed edge $(u,v)$ indicates a professor employed at institution $v$ received their PhD from institution $u$. We then algorithmically extract a continuous measure of institutional prestige using the SpringRank algorithm, which infers a linear embedding of nodes that maximizes certain predictive properties~\cite{de2018physical}. We normalize the resulting prestige scores so a unit difference between two institutions implies that, in a hiring event $(u,v)$, the PhD institution $u$ is expected to be the higher ranked institution 80\% of the time. This continuous measure allows for meaningful non-integer differences in prestige between institutions, which is valuable in our regression analysis.

\section{Subfield Inference of Publications}\label{sec:si_word_extraction} 
We use a semi-supervised topic model to assign computing publications to subfields \cite{gallagher2017anchored}. The topic model accepts lists of ``anchor words'' as priors for each topic. These words are intended to guide the topic model to delineate publication titles into groups that correspond with the research areas of computing. We use both of automated methods and domain expertise to reach an informative set of anchor words. 

We identify 2-4 publication venues (conferences and journals) dedicated specifically to each research area in our taxonomy of computing. We use publication titles indexed in DBLP from these venues (n = 232,350) as a labeled dataset. For example, the titles of publications in the Conference on Computer Vision and Pattern Recognition (CVPR) and in the the International Conference on Computer Vision (ICCV) are used as examples of titles for the research area \emph{Image Processing \& Computer Vision} (SI~Table~\ref{tab:cs_structure}). 

Using the labeled dataset, we find anchor words for each subfield that are both relatively specific to that subfield (i.e. when the anchor word is included in a title, the title belongs to a paper in the given research area at least 20\% of the time) and are measured to have high mutual information with the subfield labels in the dataset \cite{jagarlamudi2012incorporating}. Examples of words and bigrams that were algorithmically extracted as anchor words for \emph{image processing \& computer vision} are: \emph{image}, \emph{3d}, \emph{estimation}, \emph{recognition}, \emph{object}, \emph{motion}, \emph{deep}, \emph{segmentation}, \emph{video}, \emph{visual}, and \emph{learning for}. Although a perfect set of anchor words is not required for the topic model to learn a reasonable clustering of paper titles into computing research areas, we audit the algorithmically extracted words by hand to remove words that are potentially too general, and to add words that we believe to be missing. In the case of \emph{image processing \& computer vision}, we remove \emph{estimation} and \emph{learning for} from the list of anchor words for being subfield ambiguous. 

We use the audited anchor words as lexical priors in a topic modeling algorithm \cite{gallagher2017anchored}. The words inferred by the topic model to be the most informative of research area (presented in SI~Table~\ref{tab:cs_structure}) do not necessarily contain all of the anchor words that were provided as input. For the purposes of this paper, publication topic classification errors within subfield research areas are not important, as our analysis is aggregated to the subfield level. For example, if publications in \emph{interactive systems} --a research area that we believe has sub-par quality of inferred words by the topic model--are misclassified as being within \emph{computer-supported cooperative work}, this type of misclassification will not effect the interpretation of our results because the two research areas both belong to human-computer interaction on the subfield level.

\newcolumntype{C}{>{\arraybackslash}X} 

\begin{table*}
\tiny
\begin{tabularx}{\textwidth}{llC}
\toprule
                        Subfield &                               Research Area &                                                                                                                                 Words \\
\midrule
\midrule
          \rlap{\textbf{Computational Learning}} &                                          &                                                                                                                                    \\
                              &                                          AI &     learning, neural, deep, knowledge, neural networks, representation, reinforcement, reinforcement learning, reasoning, adversarial \\
                              &         Data Mining \& Information Retrieval &           mining, discovery, data mining, recommendation, discovering, data streams, recommender, commerce, frequent, recommendations \\
                              &          Image Processing \& Computer Vision &                                                      detection, image, recognition, deep, object, video, 3d, visual, images, tracking \\
                              &                            Machine Learning & stochastic, bayesian, reinforcement, reinforcement learning, kernel, gaussian, dimensional, minimization, gradient, feature selection \\
                              &                                         NLP &                                      neural, language, machine, semantic, task, text, extraction, supervised, languages, unsupervised \\
                              \midrule

      \rlap{\textbf{Human-Computer Interaction}} &                                          &                                                                                                                                    \\
                              &         Computer-Supported Cooperative Work &                       collaborative, collaboration, communities, awareness, health, teams, care, practices, participation, assessment \\
                              &                   Human-Computer Interfaces &                           interface, intelligent, reality, interfaces, user, human, augmented reality, virtual reality, gesture, gaze \\
                              &                         Interactive Systems &                                                                                  designing, can, you, are, do, that, be, we, when, it \\
                              &                            Mobile Computing &                                         wireless, devices, hoc networks, wireless sensor, radio, hoc, ad hoc, ad, communications, iot \\
                              &                            Social Computing &                                          time, real, real time, time series, series, time systems, world, narrative, real world, long \\
                              \midrule

     \rlap{\textbf{Interdisciplinary Computing}} &                                          &                                                                                                                                    \\
                              &         Computational Bio. \& Bioinformatics &                                          dynamics, protein, functional, sequence, evolution, gene, cell, expression, simulations, dna \\
                              &                     Computational Economics &                                              games, mechanisms, agent, pricing, game, multi agent, market, theoretic, auctions, price \\
                              &                Computing Education Research &                             programming, computer, science, education, teaching, students, student, computer science, special, course \\
                              &                                Data Science &                bayesian, regression, estimation, distributions, high dimensional, sample, multivariate, mixture, likelihood, interval \\
                              &             Social and Information Networks &                               networks, network, web, social networks, social, social media, recommendation, twitter, news, peer peer \\
                              \midrule

 \rlap{\textbf{Numerical \& Scientific Computing}} &                                          &                                                                                                                                    \\
                              &                      Computational Geometry &                                            geometric, convex, planar, points, shortest, plane, curves, diagrams, dimensions, spanning \\
                              &                           Computer Graphics &                                                       virtual, mesh, reality, gpu, rendering, surfaces, light, color, facial, texture \\
                              &                     Modeling and Simulation &                             simulation, scale, large, large scale, discrete event, experimental, particle, multi scale, swarm, supply \\
                              &                      Numerical Optimization &                                   problems, finite, mixed, nonlinear, schemes, convergence, equations, optimization, numerical, order \\
                              \midrule

           \rlap{\textbf{Programming Languages}} &                                          &                                                                                                                                    \\
                              &                     Language Implementation &                               compiler, flow, types, dynamic, higher order, optimizations, register, compilation, optical, collection \\
                              &                           Program Reasoning &          verification, symbolic, logic, abstraction, model checking, interpretation, verifying, reachability, proving, logic programs \\
                              \midrule

            \rlap{\textbf{Software Engineering}} &                                          &                                                                                                                                    \\
                              &                        Software Engineering &                                 software, programs, automated, code, verification, development, tool, engineering, program, execution \\
                              \midrule

                         \textbf{Systems} &                                          &                                                                                                                                    \\
                              &                       Computer Architecture &                        architecture, hardware, architectures, cache, chip, processor, memory, processors, multiprocessor, parallelism \\
                              & Comp. Networks \& Comm. Protocols &                                 networks, wireless, sensor, energy, scheduling, routing, sensor networks, service, internet, protocol \\
                              &                           Computer Security &                     security, privacy, secure, attacks, cyber, authentication, attack, access control, encryption, privacy preserving \\
                              &                                   Databases &                                           management, processing, data, database, query, databases, storage, queries, relational, big \\
                              &                       Distributed Computing &                  distributed, fault, asynchronous, broadcast, distributed systems, tolerant, consensus, clock, radio networks, mutual \\
                              &                  High-Performance Computing &                                     performance, parallel, grid, high performance, high, scientific, core, gpu, compiler, parallelism \\
                              &                           Operating Systems &                              storage, file, cloud, aware, operating, computing, flash, file system, operating system, storage systems \\
                              &                                    Robotics &                                           control, mobile, sensor, visual, robot, tracking, motion, sensing, localization, autonomous \\
                              \midrule

      \rlap{\textbf{Theory of Computer Science}} &                                          &                                                                                                                                    \\
                              &                                  Algorithms &                                   algorithms, graphs, problems, approximation, bounds, solving, minimum, nonlinear, polynomial, lower \\
                              &                                    Automata &                          cellular, allocation, automata, end, end end, resource allocation, over, finite state, congestion, bandwidth \\
                              &                    Computational Complexity &                                 complexity, lower, lower bounds, hardness, np, reductions, bound, randomness, functions, pseudorandom \\
                              &                                Cryptography &                                   secure, key, authentication, encryption, public, signatures, proofs, signature, cryptographic, hash \\
                              &                           Quantum Computing &                                                       state, codes, key, channels, circuits, quantum, signature, error, states, party \\
\bottomrule
\end{tabularx}
\caption{Subfield of structure of computing according to crowdsourced domain expertise with the top 10 words and bigrams inferred to be the most informative for partitioning publication titles into research areas by our semi-supervised topic model.} \label{tab:cs_structure}
\end{table*}

\begin{figure*}[h] 
\includegraphics[width=0.95\textwidth]{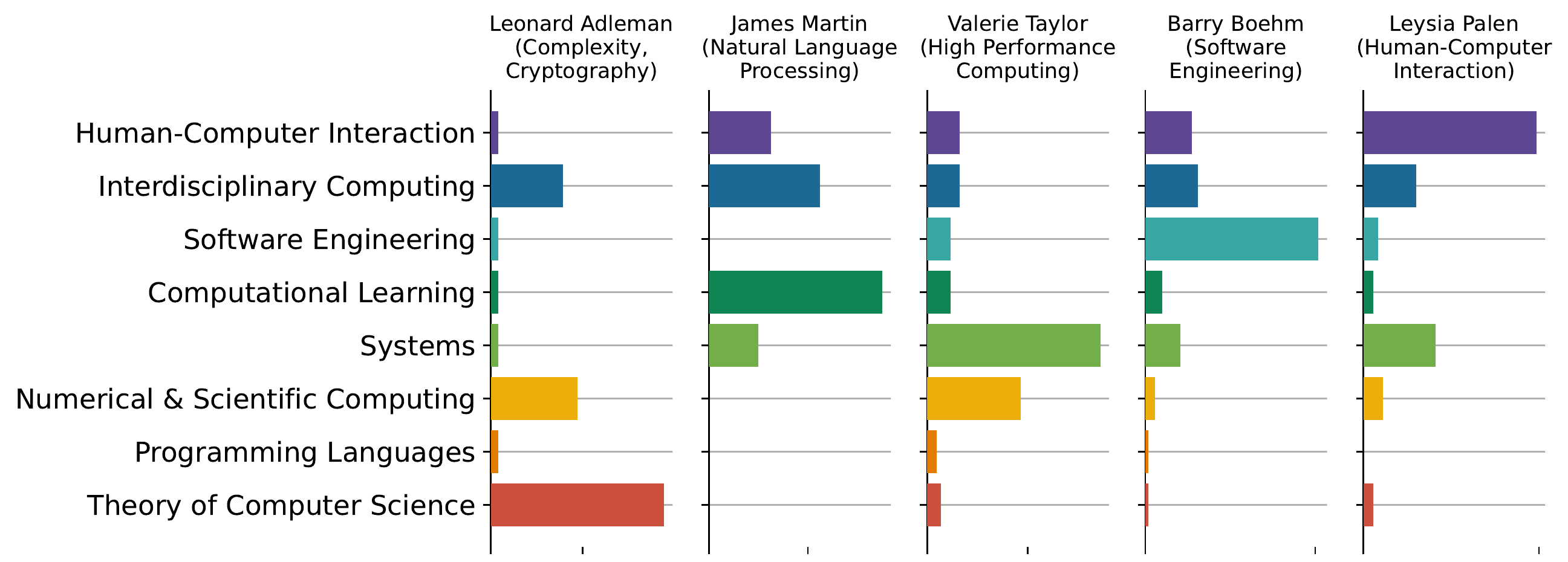} 
\caption{Inferred publication distributions over subfields for five well-known computing faculty.}\label{fig:author_distributions} 
\end{figure*}

\begin{table*} 
\begin{tabular}{l *{8}{d{3.3}}} 
\toprule 
{} & \mc{$N$ (sample)} & \mc{\% White} & \mc{\% Asian} & \mc{\% Black} & \mc{\% Hispanic} & \mc{\% Native} & \mc{\% NHPI} & \mc{\% Other} \\ 
\midrule 
Theory of Computer Science & 54.5 & 69.4 & 22.2 & 2.7 & 1.2 & 0.1 & 0.0 & 4.4 \\ 
Programming Languages & 20.5 & 71.8 & 23.4 & 0.4 & 2.4 & 0.0 & 0.1 & 1.9 \\ 
Numerical \& Scientific Computing & 46.3 & 73.5 & 20.6 & 1.5 & 1.9 & 0.1 & 0.0 & 2.3 \\ 
Systems & 143.5 & 69.8 & 24.2 & 2.3 & 1.6 & 0.2 & 0.1 & 1.9 \\ 
Computational Learning & 97.7 & 67.0 & 25.5 & 2.6 & 2.2 & 0.1 & 0.1 & 2.4 \\ 
Software Engineering & 34.5 & 74.5 & 18.5 & 2.8 & 2.6 & 0.1 & 0.1 & 1.3 \\ 
Interdisciplinary Computing & 89.8 & 72.0 & 21.4 & 2.7 & 1.5 & 0.2 & 0.0 & 2.1 \\ 
Human-Computer Interaction & 60.1 & 74.3 & 18.7 & 3.2 & 2.1 & 0.4 & 0.1 & 1.2 \\ 
\midrule 
All of Computing (total) & 547.0 & 70.8 & 22.5 & 2.5 & 1.8 & 0.2 & 0.1 & 2.2 \\ 
\midrule 
Computer Science PhDs* & - & 77.9 & 15.4 & 2.4 & 2.2 & 0.2 & - & 1.9 \\ 
U.S Population* & - & 70.0 & 3.1 & 11.1 & 10.7 & 0.8 & 0.1 & 4.3 \\ 
\bottomrule 
\end{tabular} 
\caption{Faculty racial compositions for the 8 computing subfields and two reference populations. Reference populations are weighted based on when faculty received their PhD. Native Hawaiian or other Pacific Islander is abbreviated to NHPI. } \label{tab:frac_race} 
\end{table*}

\begin{figure*}[h] 
\includegraphics[width=0.95\textwidth]{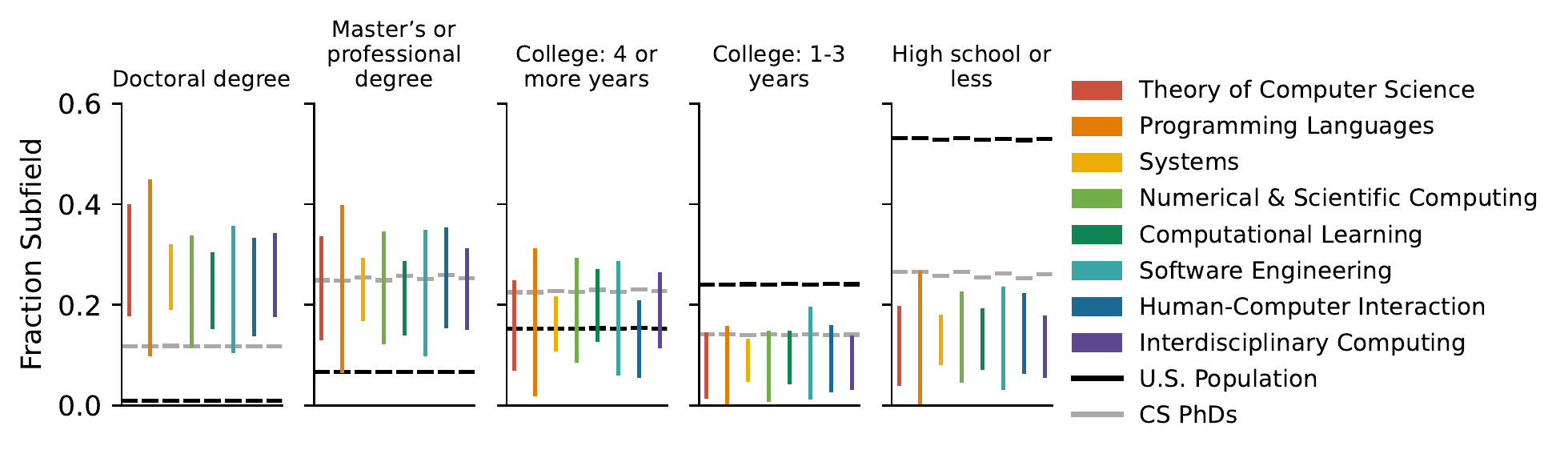} 
\caption{Representation of faculty with respect to their parents' highest level education by computing subfield, with 95\% confidence intervals, along with expected levels of representation according to two time-adjusted reference populations (see text), the U.S.~population (black lines) and computing PhDs (grey lines) \cite{census1993population, sed2020national}. Parents' highest levels of education do not differ significantly across subfields ($\chi^2 = 9.91$, $N = 570$, $p = 0.99$), but across all subfields, faculty are more likely to have parents with doctoral degrees and less likely to have parents who did not attend college than both benchmark populations.}\label{fig:parents_edu} 
\end{figure*} 

\begin{figure}[] 
\includegraphics[width=0.45\textwidth]{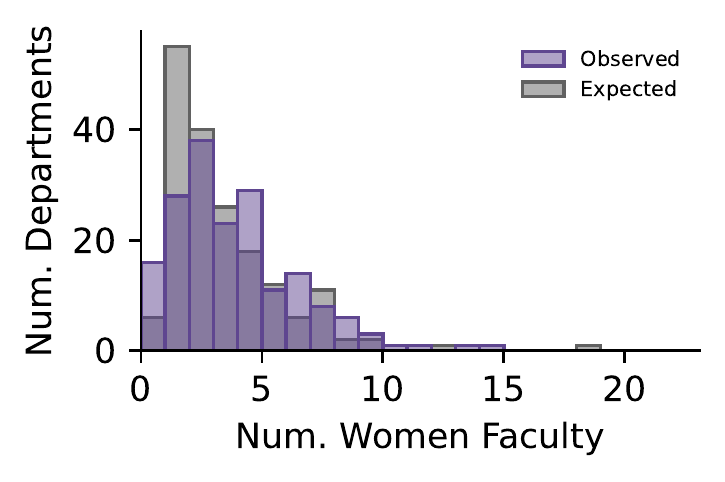} 
\caption{Histograms of the observed and expected number of women in computing departments based on a basic binomial model of hiring where each hire is equally likely to be a women. Departments are more likely to have relatively small numbers of women and relatively large numbers of women than expected based on a binomial model of hiring.}\label{fig:overdisperson} 
\end{figure} 

\section{Subfield Assignment Scheme}\label{sec:si_assignment} 

In the main text we present an analysis of computing subfields based on an assignment scheme in which faculty may belong to multiple subfields in proportion their share of publications in each subfield. This approach can capture the range of research focuses for faculty working across the boundaries of multiple subfields. 

Here, we consider an alternate assignment scheme where faculty are assigned to the subfield in which they have the most publications (modal assignment). This method of subfield assignment may miss nuance in some faculty’s research engagements, but provides an alternative view of subfields that emphasizes faculty specialties. Modal assignment also has the advantage of being especially robust to topic modeling misclassifications of individual publications. 

Some subfields are smaller in size under the modal assignment scheme, including human-computer interaction, numerical and scientific computing, and interdisciplinary computing (SI~Table~\ref{tab:si_subfield_sizes}). This finding suggests that many researchers in these fields are not as specialized within them as they are within other fields. Conversely, systems, computational learning, and software engineering are larger under the modal assignment scheme (SI~Table~\ref{tab:si_subfield_sizes}), so researchers who are modally specialized in these subfields are likely to also publish in other areas. 

The gender compositions of subfields also change modestly under modal assignment. In particular, interdisciplinary computing and human-computer interaction have higher concentrations of women faculty (SI~Table~\ref{tab:si_subfield_sizes}), which suggests that women publishing in these two fields tend to be more specialized to the fields than their male peers, who are more likely to specialize in another subfield. 

We observe more variation in subfields’ changes in relative size and gender diversity over time under the modal assignment scheme. A new computing professor under modal assignment will contribute to only one field rather than being more smoothly distributed across fields in proportion to their publications, which may partially explain the observed higher variation. 

Irrespective of choice in assignment scheme, we are lead to the same key conclusions. We find that there are significant faculty gender differences across the subfields of computing ($\chi^{2}$, $N = 4,421$, $p < 0.01$, modal assignment) and that there are prestige differences across the subfields of computing which are correlated with subfield gender composition. Particularly, faculty working in subfields with higher concentrations of women tend to be lower in the hierarchy of academic institutions, even after controlling for relevant covariates (Pearson’s $R= 0.77$, $p = 0.03$, modal assignment).

\begin{table*} 
\begin{tabular}{l *{5}{d{7.3}}} 
\toprule 
{} & \multicolumn{2}{l}{\textbf{Shared Assignment}} & \multicolumn{2}{l}{\textbf{Modal Assignment}} \\ 
& \multicolumn{1}{c}{$N$} & \multicolumn{1}{c}{\% Women}& \multicolumn{1}{c}{$N$} & \multicolumn{1}{c}{ \% Women} \\ 
\midrule 
Theory of Computer Science & 573.2 & 13.1 & 553.0 & 11.4 \\ 
Programming Languages & 181.1 & 14.2 & 135.0 & 11.5 \\ 
Numerical \& Scientific Computing & 478.8 & 14.5 & 371.0 & 15.3 \\ 
Systems & 1486.1 & 14.6 & 1718.0 & 14.2 \\ 
Computational Learning & 950.8 & 17.9 & 1059.0 & 16.6 \\ 
Software Engineering & 317.6 & 18.9 & 412.0 & 17.4 \\ 
Interdisciplinary Computing & 904.1 & 19.7 & 821.0 & 23.1 \\ 
Human-Computer Interaction & 580.4 & 20.0 & 403.0 & 23.0 \\ 
\bottomrule 
\end{tabular} 

\caption{Subfield size and fraction women faculty under two subfield assignment schemes.} \label{tab:si_subfield_sizes} 
\end{table*}

\section{Faculty Hiring Models}\label{sec:si_hiring_models} 

To test whether departments' subfield compositions are predictive of their gender compositions, we compare the likelihood of the observed data under two models of faculty hiring: a null model, where each hire is equally likely to be a women, and a subfield-informed model, where the probability that a hire is a woman varies based on a department's current subfield composition. 

The null model of faculty hiring is a simple binomial model, where the likelihood of the observed number of women in department $i$ can be expressed as 
$$\mathcal{L}_i(x_i, n_i, p) = \binom{n_i}{x_i}p^{x_i}(1-p)^{n_i-x_i}$$ 
where $x_i$ is the number of women in department $i$, $n_i$ is the total number of people in department $i$, and the parameter $p$ is the overall proportion of women across all departments (0.167, Table \ref{tab:frac_women}). We can thus express the total likelihood of this model across all departments as 
$$\mathcal{L}_{null} =\prod_{i=1}^m \binom{n_i}{x_i}p^{x_i}(1-p)^{n_i-x_i}$$ 
where $m$ is the total number of computing departments. 

The subfield-informed model takes departments' subfield compositions into consideration to determine the probabilities of hiring women faculty, and as such has 8 total parameters, $\vec{p} = [p_1, p_2, ..., p_8]$ where $p_j$ is the overall fraction of women in subfield $j$. Under this model, the expected probability that a hire is a women in a given department is calculated as a function of its distribution over subfields and the gender composition of those subfields 
$$p_i = \sum_{j=1}^8\frac{n_{ij}}{n_i}p_j$$ 
where $n_{ij}$ is the number of faculty in department $i$ conducting research in subfield $j$. Using the binomial likelihood function as before, we calculate the total likelihood of the observed data under the subfield informed model 
$$\mathcal{L}_{subfields} =\prod_{i=1}^m \binom{n_i}{x_i}p_i^{x_i}(1-p_i)^{n_i-x_i}$$ 

The observed data are more likely under the subfield informed model (log-likelihood = $-481.8$) than the null model (log-likelihood = $-488.1$). To estimate the relative quality of each model, we use the Akaike information criterion (AIC) which weighs model fit (measured by log-likelihood) against model complexity (measured by number of parameters), and we find that the null model is the preferred model for faculty hiring, due to its lower relative complexity. 

Within computing, we observe four subfields with especially low gender diversity which lag the other four subfields in gender diversity by roughly 25 years (Fig.~\ref{fig:predict}B). Because subfields within these two groups are similar in gender composition, it is possible that we overspecified our subfield informed model by providing too many parameters, and instead should only consider two parameters: the gender composition within these two groups of subfields. Taking this approach, we find that the model (log-likelihood = $-482.5$) outperforms the two pervious models under the AIC. We find the same result under the Bayesian information criterion (BIC), which penalizes complex models to a higher degree. 

Neither model fully captures the faculty hiring process: the presence of women faculty is overdispersed with respect to both the subfield-informed and null binomial models for faculty hiring, where both models under predict the number of departments with especially low and especially high numbers of women faculty (SI~Fig.~\ref{fig:overdisperson}). Contrary to the assumptions of the binomial models, this overdisperson suggests non-independence of faculty hires where a departments' probability of hiring women faculty may be related to the number of women faculty it already employs or related to its particular departmental climate, which is not considered in the binomial models. 

\section{Prestige Regression Analysis}\label{sec:si_regression} 

To estimate prestige differences across the computing subfields controlling for faculty degree prestige, publication productivity, and gender, we conduct a multiple weighted least squares (WLS) regression. WLS is used over standard multiple linear regression (MLR) in order to account for model heteroskedasticity, which arises in the form of higher residual variance for higher predicted prestige outcomes. This pattern in the residuals is expected due to the strongly hierarchical nature of faculty hiring, where the most prestigious institutions train a large share of the field’s faculty \cite{clauset2015systematic, way2016gender}. 

To find the appropriate weights for each point, we first train a MLR model with faculty prestige as the outcome variable and all predictors of interest as explanatory variables. The absolute value of this model’s residuals are used as the outcome variable in a subsequent MLR model, with estimated values $\hat{y}_i$ related to the WLS weights in the following relationship: $$w_i=\frac{1}{\hat{y}_i^2}$$The goal of WLS is to give less weight to points that are expected to have high variance about the mean, so we use the squared inverse of these predictions as weights in the final WLS model. In this case, the WLS model has no major qualitative differences from the corresponding MLR model, but we present results for the WLS regression as its parameter estimations are best linear unbiased estimators. Regression results for the analysis are included in SI~Table~\ref{tab:regression}. 

\begin{table*} 
\begin{tabular}{l *{5}{d{7.3}}} 
\toprule 
& \multicolumn{2}{c}{\textbf{Model 0}} & \multicolumn{2}{c}{\textbf{Model 1}} \\ 
\cline{2-5} 
{} & \multicolumn{1}{c}{Coef.} & \multicolumn{1}{c}{Rank Difference} & \multicolumn{1}{c}{Coef.} & \multicolumn{1}{c}{Rank Difference} \\ 
\midrule 
\midrule 
\textbf{Subfields} & & & & \\ 
\midrule 
\quad Ref. = Human-Computer Interaction & - & - & - & - \\ 
\quad \rule{0pt}{4ex}Interdisciplinary Computing & 0.21 & 5 & 0.11 & 3 \\ 
\quad & (0.17) & & (0.16) & \\ 
\quad \rule{0pt}{4ex}Software Engineering & 0.14 & 3 & 0.34 & 8 \\ 
\quad & (0.24) & & (0.19) & \\ 
\quad \rule{0pt}{4ex}Computational Learning & 0.72** & 17 & 0.23 & 5 \\ 
\quad & (0.15) & & (0.13) & \\ 
\quad \rule{0pt}{4ex}Systems & 0.82** & 19 & 0.43** & 10 \\ 
\quad & (0.12) & & (0.11) & \\ 
\quad \rule{0pt}{4ex}Numerical \& Scientific Computing & 1.07** & 26 & 0.60** & 14 \\ 
\quad & (0.20) & & (0.18) & \\ 
\quad \rule{0pt}{4ex}Programming Languages & 1.53** & 37 & 0.51 * & 12 \\ 
\quad & (0.28) & & (0.25) & \\ 
\quad \rule{0pt}{4ex}Theory of Computer Science & 1.43** & 36 & 0.42** & 10 \\ 
\quad & (0.16) & & (0.14) & \\

\textbf{Controls} & & & & \\ 
\midrule 
\quad Degree Prestige (SpringRank) & - & - & 0.61** & 14 \\ 
\quad & & & (0.13) & \\ 
\quad \rule{0pt}{4ex}Publication Productivity (Z-score) & - & - & 0.21** & 5 \\ 
\quad & & & (0.03) & \\ 
\quad \rule{0pt}{4ex}Women & - & - & -0.11 & -3 \\ 
\quad & & & (0.07) & \\ 
\bottomrule 
\tiny{Standard errors in parentheses. * p < 0.05; ** p < 0.01.} & & \\ 
\end{tabular} 
\caption{ Regression results for prestige regression models with faculty departmental rank as the outcome variable. Model 0 reports multiple linear regression results with only subfield participation variables as predictors. Model 1 reports multiple weighted least squares results with both subfield participation variables and control variables. Model coefficient estimates are reported in units of SpringRank, and the expected rank differences in integer units of computing departments associated with each SpringRank coefficient are additionally reported for interpretability.}\label{tab:regression} 
\end{table*} 

\begin{figure*}[t] 
\includegraphics[width=0.60\textwidth]{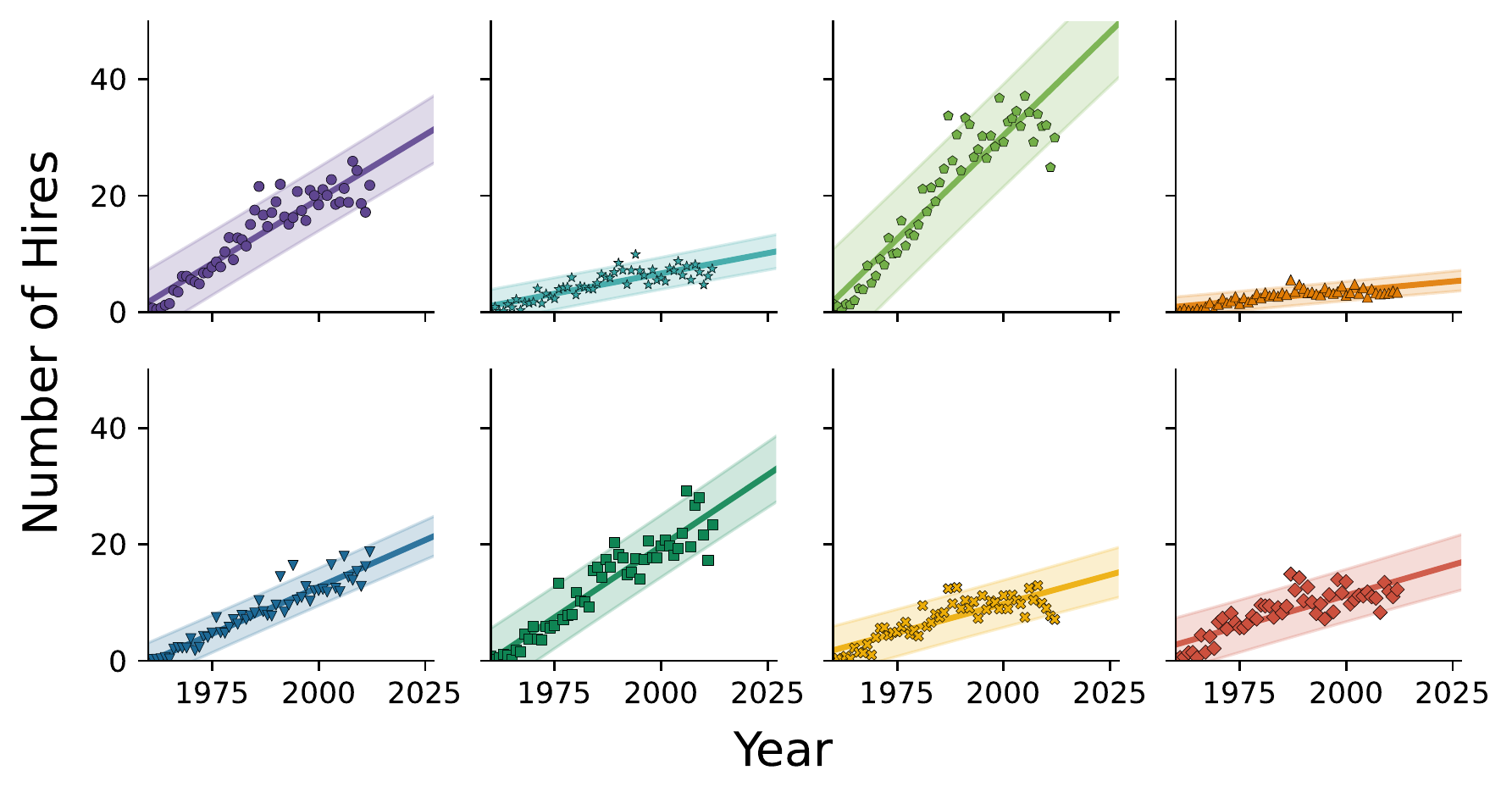} 
\includegraphics[width=0.35\textwidth]{predict_legend_binomial.pdf} 
\caption{Number of hires observed in our dataset for each subfield from 1960-2018. Linear regression trend lines shown with 95\% prediction intervals.}\label{fig:predict_num_hires} 
\end{figure*} 

\begin{figure*}[t] 
\includegraphics[width=0.60\textwidth]{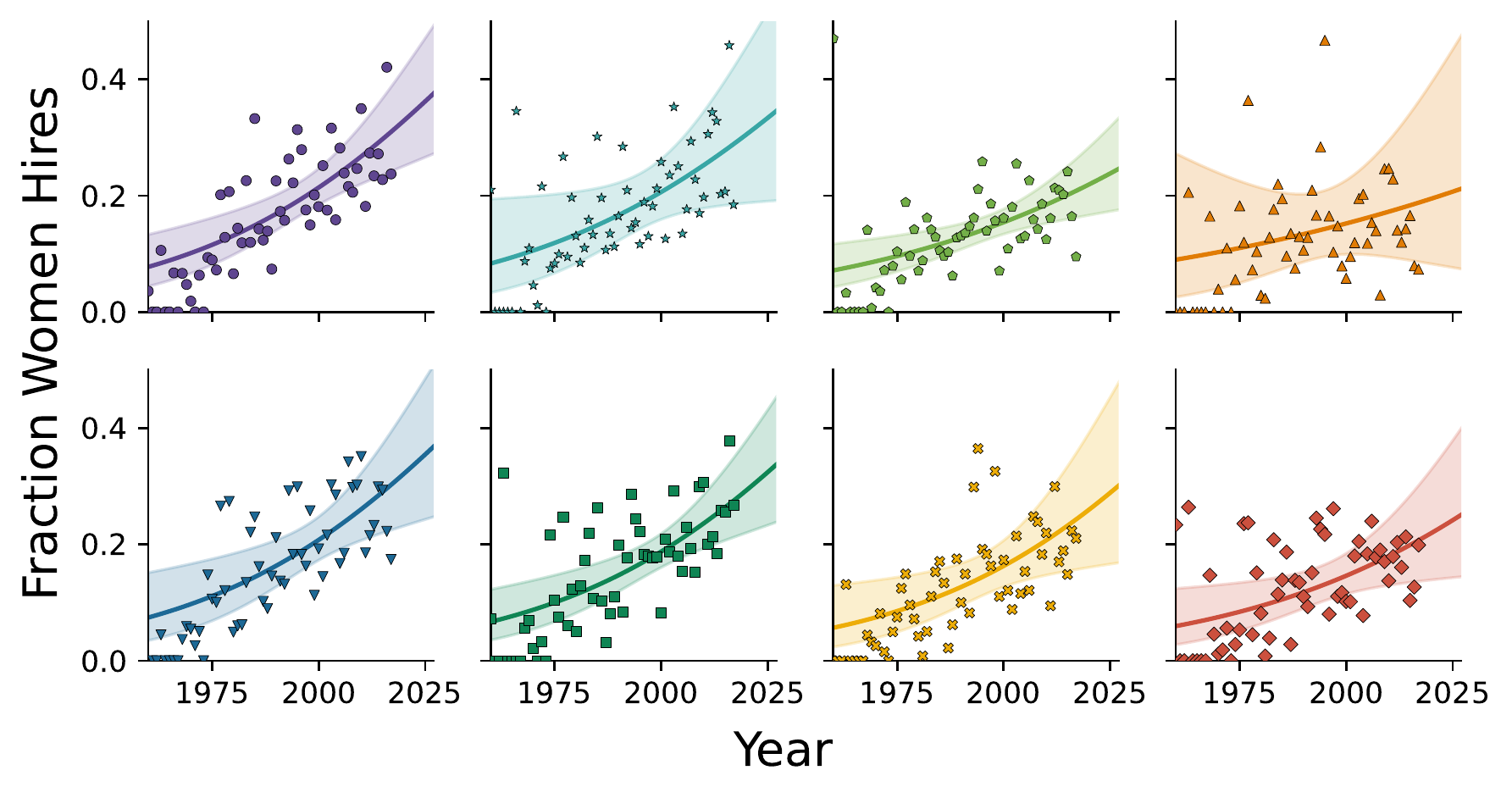} 
\includegraphics[width=0.35\textwidth]{predict_legend_binomial.pdf} 
\caption{Fraction yearly women hires in our dataset for each subfield from 1960-2018. Trend lines fitted to data by univariate binomial regression.}\label{fig:predict_binomial_hiring} 
\end{figure*}

\end{document}